\documentclass[a4paper,american,aps,citeautoscript,floatfix,longbibliography,pdftex,pra,
superscriptaddress,showpacs,preprint]{revtex4-1}

\usepackage{adjustbox}

\usepackage{enumerate,tensor}
\usepackage[normalem]{ulem}
\usepackage{natbib} 
\usepackage{hyperref}
\usepackage{amsmath,amssymb,amsmath}
\usepackage{graphicx}
\usepackage{dcolumn}
\usepackage{bm}
\usepackage{color}

\makeatletter
\def\paragraph{\@startsection{paragraph}{4}{10pt}{-1.25ex plus -1ex minus -.1ex}{0ex plus 0ex}{\normalsize\textit}}
\renewcommand\@biblabel[1]{#1}
\renewcommand\@makefntext[1]%
{\noindent\makebox[0pt][r]{\@thefnmark\,}#1}
\DeclareRobustCommand\onlinecite{\@onlinecite}
\def\@onlinecite#1{\begingroup\let\@cite\NAT@citenum\citealp{#1}\endgroup}
\def\tagform@#1{\maketag@@@{\ignorespaces#1\unskip\@@italiccorr}}
\let\orgtheequation\theequation
\def\theequation{(\orgtheequation)}
\makeatother

\begin{document}

\author{C. Chandre}
\affiliation{Aix Marseille Univ, CNRS, Centrale Marseille, I2M, Marseille, France}

\author{Jorge Mahecha}
\affiliation{Instituto de F\'{\i}sica, Universidad de Antioquia UdeA; 
Calle 70 No. 52-21, Medell\'{\i}n, Colombia}
  
\author{J. Pablo Salas}
\affiliation{\'Area de F\'{\i}sica, Universidad de la Rioja, 26006 Logro\~no, La Rioja, Spain}

\title{Driving the formation of the RbCs dimer by a laser pulse. A nonlinear dynamics approach} 

\date{\today}

\begin{abstract}
We study the formation of the RbCs molecule by an intense laser pulse using nonlinear dynamics.
Under the Born-Oppenheimer approximation, the system is modeled by a two degree of
freedom rovibrational Hamiltonian, which includes
the ground electronic potential energy curve of the diatomic molecule
and the interaction of the molecular polarizability
with the electric field of the laser. 
As the laser intensity increases, we observe that the
formation probability first increases and then decreases after reaching a maximum. We show that
the analysis can be simplified to the investigation of the long-range interaction between the two atoms.  
We conclude that the formation is due to a very small change in the radial momentum of the dimer
induced by the laser pulse. From this observation, we build a reduced one dimensional
model which allows us to derive
an approximate expression of the formation probability as a function of the laser intensity. 
\end{abstract}

\maketitle 

\section{Introduction}
During the last two decades, the development of sophisticated experimental
techniques allowed one to use ultracold atoms to create two new states of
matter that can be manipulated with high precision: The Bose-Einstein condensates
(BECs)~\cite{bec1,bec2,bec3} and the Degenerate Fermi gases (DFGs)~\cite{gf1,gf2,gf3}.
Using the deep experimental background obtained with the investigations on
BEC and on DFG,  efforts have been dedicated to achieving a similar degree of control in molecular gases.
Indeed, the production and manipulation
of dense gases of cold and ultracold
molecules constitute nowadays an
active research field in Atomic and Molecular Physics.
In particular, starting from a gas of ultracold atoms, the photoassociation~\cite{A859,A819}, the
magneto-association~\cite{A855} 
and the stimulated Raman adiabatic passage 
(STIRAP)~\cite{kerman2004}, are among the usual techniques to create
cold and ultracold molecules.
These experimental techniques have been successfully applied to
form different homonuclear and heteronuclear alkali diatomic molecules
in the rovibrational ground state, such as
C$_2$~\cite{dulieu2008,dulieu2010}, LiCs~\cite{kraft}, KRb~\cite{ni}
or RbCs~\cite{dulieu2011,take,koppi}.
Furthermore, a number of
 theoretical studies have guided and promoted many of the experimental achievements.
 Among others theoretical studies, we refer the reader to Refs.\cite{dulieu2006,juarros,A861}
 and references therein. For a review about science, 
technology and applications of cold and ultracold molecules, we refer to Ref.~\cite{A854}.

All the aforementioned techniques to create molecular bound states are based on
the external control of the
interactions of atoms and molecules with electromagnetic fields. From a classical point
of view, it is of particular interest to study how the mechanical forces exerted light on 
atoms and molecules perturb their motion. Moreover, the nonlinear nature of these forces
make these systems very
appealing for classical studies because, by the external control of the strengths of
the interactions, we have at hand the possibility to
tune the system through different classical regimes.
It is worth noting at this point that the use of classical mechanics to study
microscopic systems is not new: Over the last three
decades, a plethora of studies related to the classical dynamics of atoms
and molecules in external fields can be found in the literature. Some examples of such as studies
can be found in Refs.~\cite{A19,A33,A50,A205,PRA2007,EJPD2007,A286,PLA2010,A635,A741,A714}.
Furthermore, classical studies in microscopic systems have revealed
themselves as a power tool to understand quantum mechanical results in many cases (see e.g.~\cite{A19,gutzwiller,A311,A585,A714,Blumel} and references therein).

Here we use nonlinear dynamics to explore
the feasibility of creating cold
diatomic molecules by using a
strong linearly polarized laser pulse.
While the usual techniques to create cold and ultracold diatomic molecules require the use of several excited
electronic states, we describe here how the nonlinear mechanical force exerted by a laser field
on an initially unbounded pair of cold atoms in their ground electronic state can lead to the formation
of a bounded dimer. More precisely, we focus on the influence of the
laser field  in the formation of RbCs molecules.
Besides the kinetic terms, the rovibrational Hamiltonian of the system includes two fundamental terms: namely,
the potential energy curve between the Rb and Cs atoms and the interaction between the
molecular polarizability and the laser field. Because the laser pulse contains an envelope
with ramp-up, plateau and ramp-down, the system depends explicitly on time and the
corresponding Hamiltonian has 3+1$/$2 degrees of freedom. However, by using spherical coordinates, the
number of degrees of freedom can be reduced to 2+1$/$2.
For a convenient ensemble of initial conditions, we
compute the formation probability as a function of the laser field strength, for different
values of the parameters of the pulse. In all cases we find that, as the field strength
increases from zero, the
formation probability first increases before reaching a maximum, and then decreases
for larger values of the field strength. It is worth noting that a similar behavior has been found in the
ionization probability of atoms in the presence of an intense laser field~\cite{A801,Grobe}.
From a detailed exploration of the dynamics of the system after the ramp-up, plateau and ramp-down sequences
of the laser pulse, we infer that the study of the formation mechanism can be reduced to the
investigation of the long-range interaction between the two atoms.  
Indeed, we show that the formation is due to a very small change in the radial momentum of the dimer
induced by the laser pulse. These observations allow us to build a simplified one-dimensional
Hamiltonian where only the long range terms of the potential energy curve and the molecular
polarizabilities are taken into account. From this simplified Hamiltonian, we obtain
an analytic approximate expression for the formation probability as a function of the laser intensity.
This analytic expression mimics very accurately the described behavior of the formation probability.

The paper is organized as follows: In Sec.~II we present the main ingredients of the Hamiltonian
of the system. In Sec.~III we compute the formation probability as a function
of the laser field strength. In order to get some insights into the behavior of the formation
probability, we study the particular role played by the ramp-up, the plateau and
the ramp-down of the laser pulse. The results of Sec.~III allow us to define in Sec.~IV
 a one-dimensional version of the 
full Hamiltonian which captures the main characteristics of the system. In 
Sec.~V we define the simplified Hamiltonian with only the long-range terms
of the potential energy curve and the molecular polarizabilities. We
show that this asymptotic Hamiltonian is sufficient to describe the behavior of the
formation probability. Furthermore, we  construct an analytic
expression for the formation probability which includes the parameters of the laser pulse
and the long-range parameters of the potential energy curve and the molecular polarizabilities.

\section{The Hamiltonian of the system}
Within the Born-Oppenheimer approximation, we describe the dynamics of  the
RbCs molecule
in its $^1\Sigma^+$ electronic ground state in the presence of a strong linearly polarized laser field.
The  electric field of the laser is assumed to propagate in the parallel direction of the
$z$-axis of an inertial reference frame with the origin at the center of mass of the
nuclei. For a nonresonant laser field, the Hamiltonian
of the system can be expressed as  \cite{A509}
\begin{equation}
\label{hami}
{\cal H}= \frac{P_R^2}{2\mu}+\frac{P_\theta^2}{2\mu R^2} +  \frac{P_\phi^2}{2\mu R^2\sin^2\theta} +
V(R,\theta,t),
\end{equation}
\begin{figure}[t]
\centerline{\includegraphics[scale=.9]{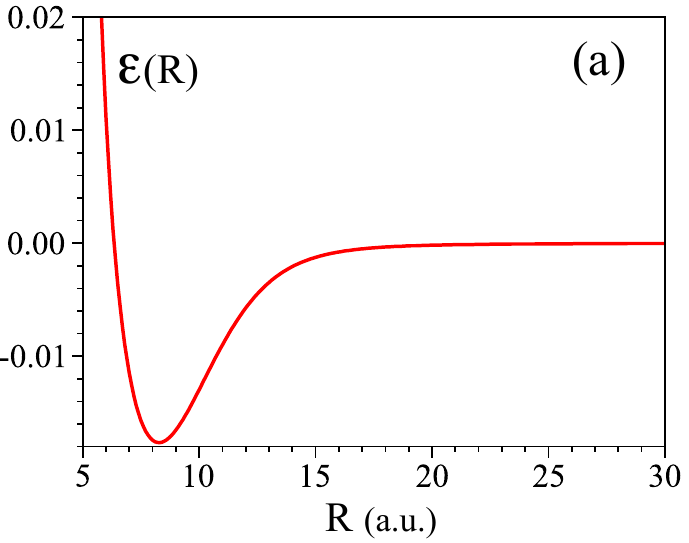} \quad 
\includegraphics[scale=.9]{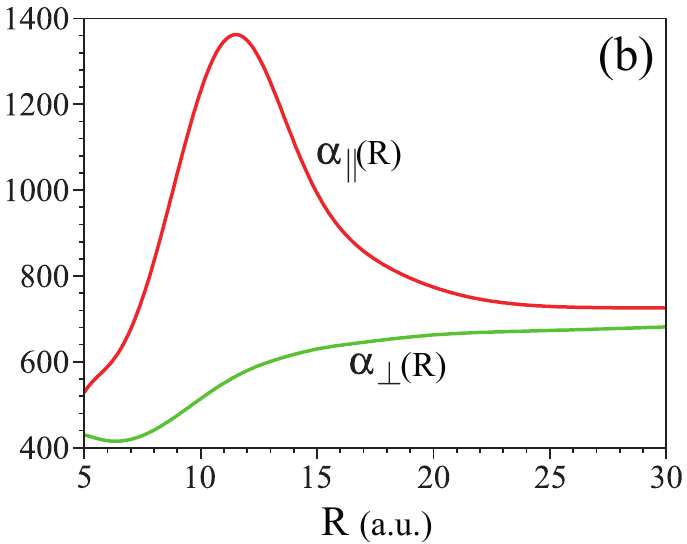}} \caption{a) Electronic potential energy curve $ \varepsilon(R)$
of the RbCs and b) parallel $\alpha_\parallel(R)$ and perpendicular
$\alpha_\bot(R)$ components of the molecular polarizability  of the RbCs molecule.}
\label{fi:curvas}
\end{figure}
\noindent
where $\mu$ is the reduced mass of the nuclei, ($R, \theta$, $\phi$) are the internuclear distance and
the Euler angles, and $(P_R, P_\theta, P_\phi)$ are the
corresponding canonically conjugate momenta. $V(R,\theta,t)$ is the potential energy surface given by
\begin{equation}
\label{efectivo}
 V(R,\theta)= \varepsilon(R)+ V_L(R,\theta,t),
\end{equation}
\noindent
which is made of the field-free adiabatic electronic potential energy curve  $\varepsilon(R)$
and the laser-molecule interaction potential $V_L(R, \theta,t)$,
\begin{equation}
\label{laser}
 V_L(R,\theta,t)= -g(t) \frac{F^2}{4}[\alpha_\parallel(R) \cos^2\theta  + \alpha_\bot(R) \sin^2\theta].
\end{equation}
\noindent
The function $g(t)$ is the laser pulse envelope and $F$ is the strength of the
electric field of the laser.  The functions $\alpha_{\parallel,\bot}(R)$
are the parallel and the perpendicular molecular polarizabilities \cite{polarizabilidad}.
The pulse envelope $g(t)$ contains a ramp-up, a plateau and a ramp-down with durations
$T_{\rm ru}$, $T_{\rm p}$ and $T_{\rm rd}$, respectively, and its profile
is taken to be \cite{A182}
\begin{equation}
g(t)=\left\{ \begin{array}{ll} \displaystyle \sin^2 \left(\frac{\pi t}{2T_{\rm ru}}\right) & 
\mbox{if } 0\leq t<T_{\rm ru},\\ 
1 & \mbox{if } T_{\rm ru}\leq t<T_{\rm ru}+T_{\rm p},\\ 
 \displaystyle \sin^2\left(\frac{\pi (t-T_{\rm ru}-T_{\rm p}-T_{\rm rd})}{2T_{\rm rd}} \right) & 
 \mbox{if } T_{\rm ru}+T_{\rm p}\leq t<T_{\rm ru}+T_{\rm p}+T_{\rm rd},\\
 0 & \mbox{elsewhere}. 
\end{array}\right.
\label{eq:pulse}
\end{equation}
This field envelope describes accurately experimental laser pulses \cite{A747}.

\medskip
In order to manage an analytical representation for the potential energy surface $V(R,\theta,t)$ for
the RbCs molecule, we have fitted the available data of $\varepsilon(R)$ \cite{pec} and
$\alpha_{\parallel,\bot}(R)$  \cite{polarizabilidad}
to three appropriate functional forms. In the case of $\varepsilon(R)$, the
fitting function of the {\sl ab initio} data includes the long-range behavior of the energy
curve which is expressed as \cite{marinescu}
\begin{equation}
\label{pecLR}
\varepsilon_{LR}(R) = -\frac{b_6}{R^6} - \frac{b_8}{R^8} - \frac{b_{10}}{R^{10}}.
\end{equation}
For the $^1\Sigma^+$ RbCs these
coefficients can be found in the literature \cite{marinescu} and their values are
reported in Table~\ref{ta:tabla1}.
The asymptotic behavior of the polarizabilities $\alpha_{\parallel,\bot}(R)$
is well described by the Silberstein expressions
\cite{Silberstein,jensen,Silberstein2}
\begin{eqnarray}
\label{Silberstein}
\alpha_\parallel^{LR}(R) &=& \frac{\alpha_{\rm RbCs} + 4 \alpha_{\rm Rb} \alpha_{\rm Cs}/R^3}
{1 - 4 \alpha_{\rm Rb} \alpha_{\rm Cs}/R^6},\nonumber\\
& & \\
\alpha_\bot^{LR}(R) &=& \frac{\alpha_{\rm RbCs} - 2 \alpha_{\rm Rb} \alpha_{\rm Cs}/R^3}
{1 -  \alpha_{\rm Rb} \alpha_{\rm Cs}/R^6}\nonumber,
\end{eqnarray}
\noindent
where $\alpha_{\rm Rb}\approx313$ a.u. and $\alpha_{\rm Cs}\approx394$ a.u.
are the atomic polarizabilities of the atoms and $\alpha_{\rm RbCs}=\alpha_{\rm Rb} + \alpha_{\rm Cs}$.
The two Silberstein expressions \ref{Silberstein}
diverge when
$R \rightarrow (4 \alpha_{\rm Rb} \alpha_{\rm Cs})^{1/6} \approx 8.8889$ a.u. 
and $R \rightarrow (\alpha_{\rm Rb} \alpha_{\rm Cs})^{1/6} \approx 7.0552$ a.u., respectively. This is
a drawback for classical calculations.
Taking into account that  computational data for the molecular polarizabilities are available up to the 
intermolecular distance of $R=30$ a.u., instead of using the analytical expression
\ref{Silberstein} to model the long-range behavior 
of $\alpha_{\parallel, \bot}$, we
append to the computational data of the molecular polarizabilities,
values of $\alpha_{\parallel, \bot}$ evaluated for $R>30$ a.u. at the Silberstein expressions
\ref{Silberstein}. This allows us to fit
the polarizabilities $\alpha_{\parallel, \bot}$ with smooth functions which
are very convenient for classical calculations.
The long-range fittings for $\alpha_{\parallel,\bot}(R)$ are given by
\begin{eqnarray}
\label{polaPerLR}
\alpha_{\bot}^{LR}(R)&=& \alpha_{\rm RbCs} + \frac{c_2}{R^2} +  \frac{c_3}{R^3} +\frac{c_4}{R^4},\\[2ex]
\label{polaParaLR}
\alpha_{\parallel}^{LR}(R)&=&\alpha_{\rm RbCs} + \frac{d_2}{R^2} +  \frac{d_3}{R^3} +\frac{d_4}{R^4}.
\end{eqnarray}
The fitting parameters $b_i$, $c_i$ and $d_i$ are shown in Table~\ref{ta:tabla1}.
The fitted curves $\varepsilon(R)$ and $\alpha_{\parallel,\bot}(R)$
are plotted in Fig. \ref{fi:curvas}.

\begin{table*}
\caption{Values of the fitting parameter for the long-range behavior of the
potential energy curve $\varepsilon(R)$ and the parallel and perpendicular
polarizabilities $\alpha_{\parallel,\bot}(R)$. All parameters are given in atomic units.}

\centering
\begin{tabular}{*{4}{lcll}|}
\hline\noalign{\smallskip}
\noalign{\smallskip}\hline\noalign{\smallskip}
$b_6 = 5284$ & \quad $b_8=730520$ & \quad $b_{10}=1.0831 \times 10^8$ \\[2ex]
$c_2=1888.9$ & \quad $c_3=-351865.9$ & \quad $c_4=1.5056\times 10^6$\\[2ex]
$d_2=1277.8$ & \quad $d_3=374596.4$ & \quad $d_4=2.7868\times 10^6$ \\
\noalign{\smallskip}\hline
\end{tabular}
\label{ta:tabla1}
\end{table*}

\begin{figure}[t]
\centerline{\includegraphics[scale=.35]{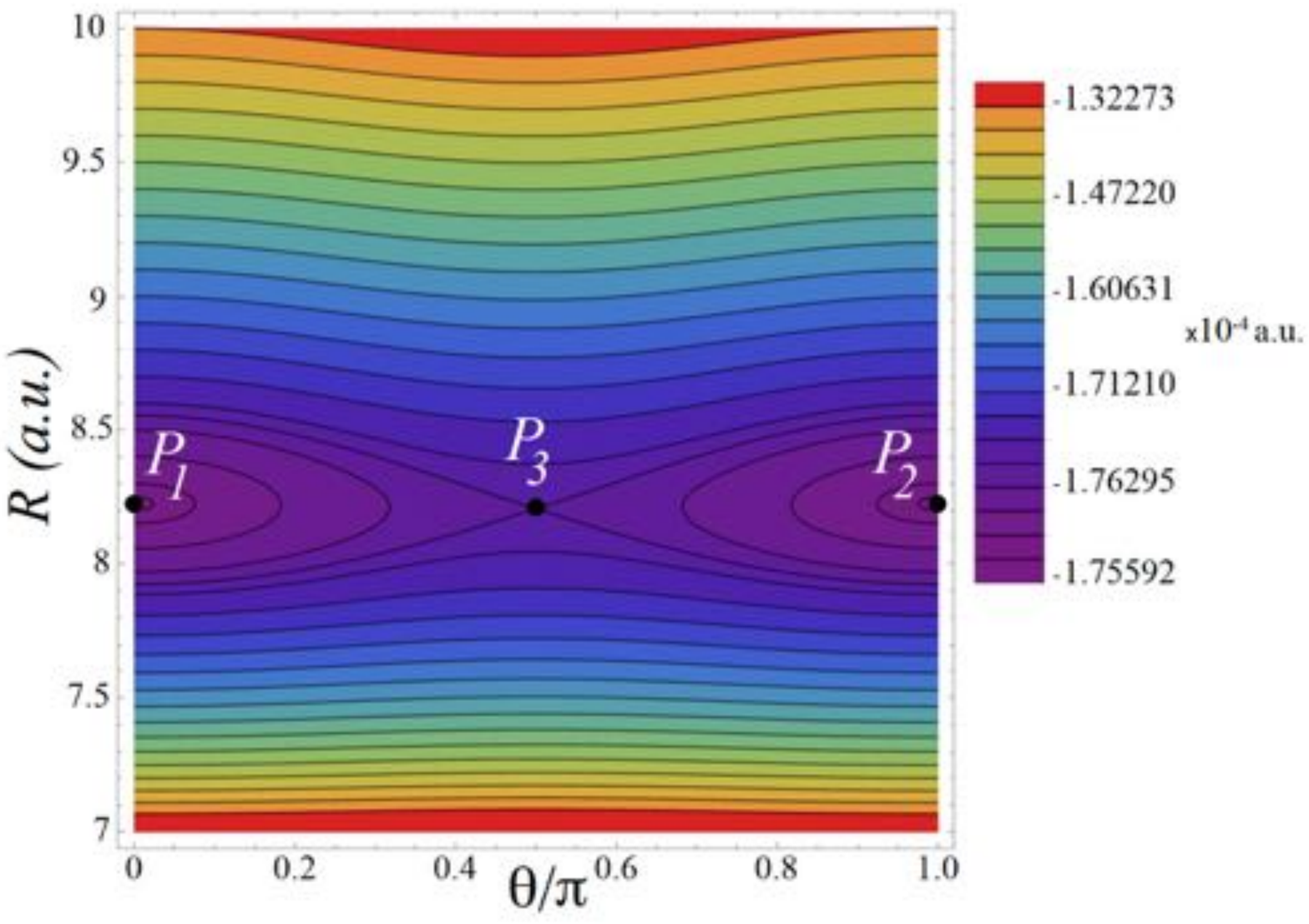}}

\centerline{ \includegraphics[scale=.35]{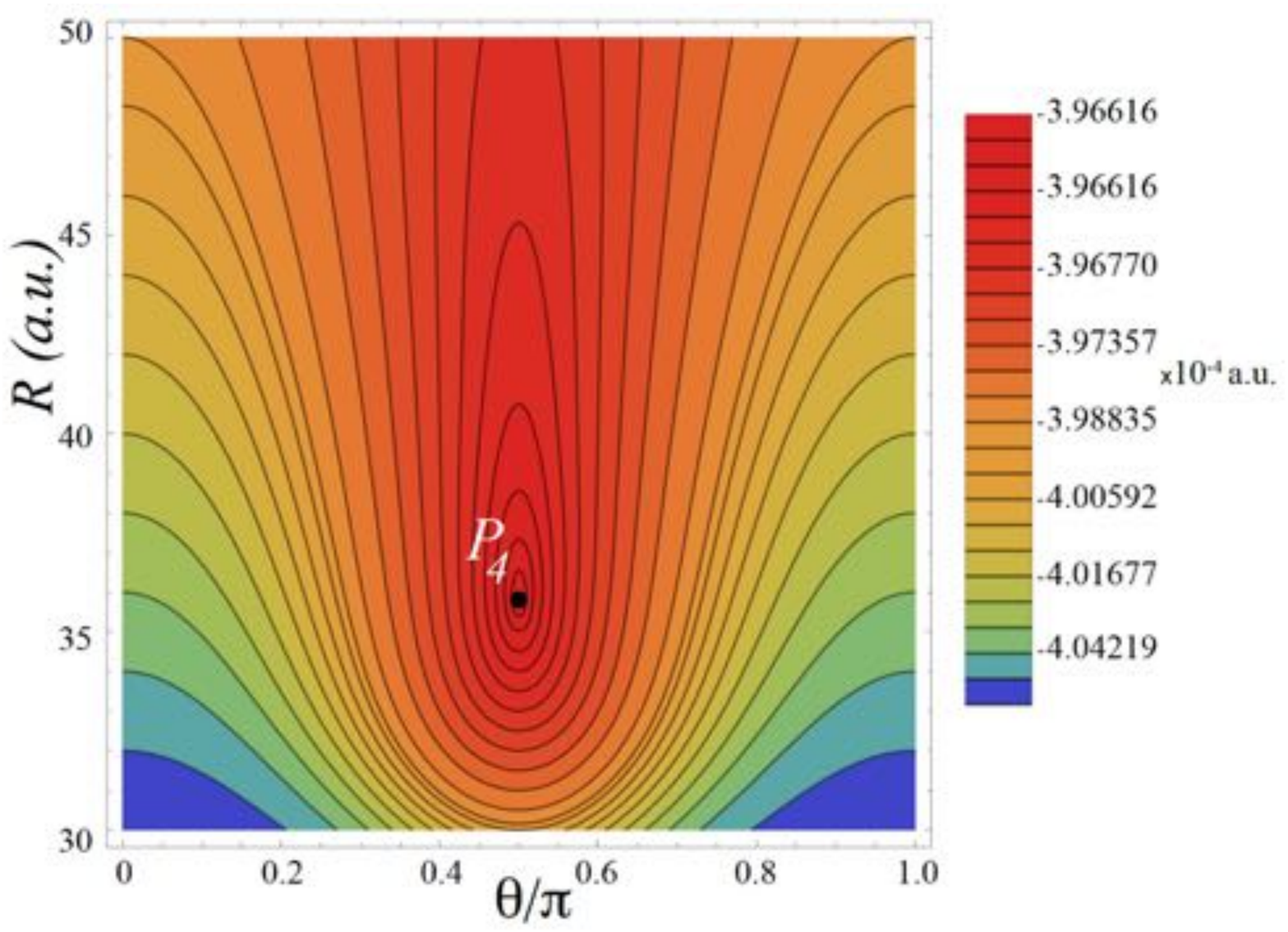}}
\caption{Equipotential curves of the potential energy
surface $V(R,\theta,t)$  during the plateau ($g(t)=1$)
for a laser field strength $F=1.5 \times 10^{-3}$ a.u.}
\label{fi:potencial}
\end{figure}

\medskip
Owing to the continuous axial symmetry of the system, the polar angle $\phi$ is cyclic in
Hamiltonian~ \ref{hami} and the z-component $P_{\phi}$ of the angular momentum is conserved.
This allows one to consider the expression  \ref{hami} as a classical Hamiltonian system
with 2+1$/$2 degrees of freedom in $(R, \theta )$.
The $1/2$ degree of freedom is due to the explicit time-dependence in ${\cal H}$.
The present study is restricted to the $P_\phi = 0$ case, i.e., the
corresponding magnetic quantum number is zero, being this particular value widely
used is several studies \cite{A181,A229}.
The landscape of the potential energy surface $V(R,\theta,t)$ during the plateau ($g(t)=1$) is
strongly determined by the polarizability. Indeed, as we can observe in Fig.\ref{fi:potencial}, for
$F=1.5 \times 10^{-3}$ a.u., the
energy surface $V(R,\theta,t)$ presents four critical points: two equivalent
minima $P_{1,2}$ at $\theta=0, \pi$ respectively, a
saddle point $P_3$ at $\theta =\pi/2$ and a maximum $P_4$ at $\theta =\pi/2$.
These critical points create two different regions of motion. 
When the energy of the molecule is below the energy of the saddle point $P_3$, the rovibrational motion of the
dimer is made of pendular states \cite{A132} around
the minima $P_{1,2}$  because the molecule is confined in one of the
potential wells  around $P_{1,2}$. In other words, we find the expected behavior
of a dimer aligned in the $\theta=0, \pi$
directions \cite{barker}.
On the other hand, when the energy of the system is above the saddle point energy, the molecule
can describe complete rotations.
Due to the ``energy hill"  around the maximum $P_4$ created by the polarizability, the
molecular bond $R$ always reaches
its largest values along the $\theta=0, \pi$ directions.
As the electric field strength $F$ increases, the maximum $P_4$ approaches the
saddle point $P_3$ and its energy
increases.
The directions $\theta=0, \pi$ together with the threshold
dissociation conditions $R \rightarrow \infty$,
$P_R  \rightarrow 0$ and  $P_{\theta} \rightarrow 0$, allow us to  get an analytical estimate of 
the dissociation energy $E_d$.
Under the condition $R \rightarrow \infty$, the function
$\varepsilon(R)$ tends to
$0$, and $\alpha_{\parallel}(\infty) = \alpha_{\perp}(\infty)=\alpha_{\rm Rb}+\alpha_{\rm Cs}$.
Then, the  approximate
value for the dissociation energy is given by
\begin{equation}
\label{eq:enegydisociacion} 
E_d \approx - \frac{F^2}{4} \ \alpha_{\parallel}(\infty) = -
\frac{F^2}{4} \ (\alpha_{\rm Rb}+\alpha_{\rm Cs}).
\end{equation}
\noindent
Thus, the molecular polarizabilities lead to a decrease of the dissociation energy to a
negative value, which depends on the electric field strength $F$ as well as on the 
polarizabilities of the atoms.
%

\section{Driving the formation of the dimer. Numerical experiments}
\label{driving}
We use Hamiltonian \ref{hami} to study the impact of the laser field
in the creation of bound molecular states. In particular, we compute numerically the formation
probability $P(F)$ as a function of the field strength $F$.
To do that, we consider a large ensemble of initially free pairs of Rb-Cs atoms, whose
dynamics is governed by the ``free" Hamiltonian
\begin{equation}
\label{hamifree}
{\cal H}_0 = \frac{P_R^2}{2\mu}+\frac{P_\theta^2}{2\mu R^2} +\varepsilon(R).
\end{equation}
\noindent
All the initial conditions $(R_0, P_R^0, \theta_0, P_{\theta}^0)$
of the ensemble 
have the same positive energy ${\cal H}_0=E_0=3\times 10^{-9}$ a.u.
 This energy roughly
corresponds to the temperature $T=1$ mK of a sample of cold
atoms in a typical photoassociation experiment \cite{A859,A331}.
The choice of the
initial states is an important issue as it is shown later on. Here $P_\theta^0$ is taken to be zero, $\theta_0$
is chosen randomly in $[0,\pi]$, and $R_0$ is chosen in
the interval $[R_{\rm min},R_{\rm max}]=[6.2319, 100]$ a.u., where 
$R_{\rm min}$ is the 
(inner) turning point of the phase trajectory of Hamiltonian \ref{hamifree} for
$P_{\theta}^0=0$.
First, let us compute the time evolution of the (unbound)
trajectory of energy $E_0=3\times10^{-9}$ a.u. starting at the initial internuclear distance
$R_0 = R_{\rm max}$ and with the inward initial radial momentum $P_R^0 \approx -0.04$ a.u. given by
Eq.~\ref{hamifree}.  We consider this orbit until it reaches again $R_{\rm max}$
with $P_R\approx 0.04$ a.u.
When the intermolecular distance $R(t)$ of this trajectory is mapped at equal time intervals, we observe that
large values of $R(t)$ are rapidly reached. In other words,  the initial conditions with
large values of $R_0$ are more likely than initial conditions with small values of $R_0$.
In this way, in order to mimic more accurately the initial states of the
system, we choose the initial conditions $(R_0, P_R^0)$ along the phase curve
\ref{hamifree} for $E_0$ at equal time steps.
It is worth noting that, with these initial conditions uniformly 
distributed over time, less than a $1\%$
of the initial conditions have values of $R_0<25$ a.u.

\begin{figure}
 \includegraphics[width=0.5\textwidth]{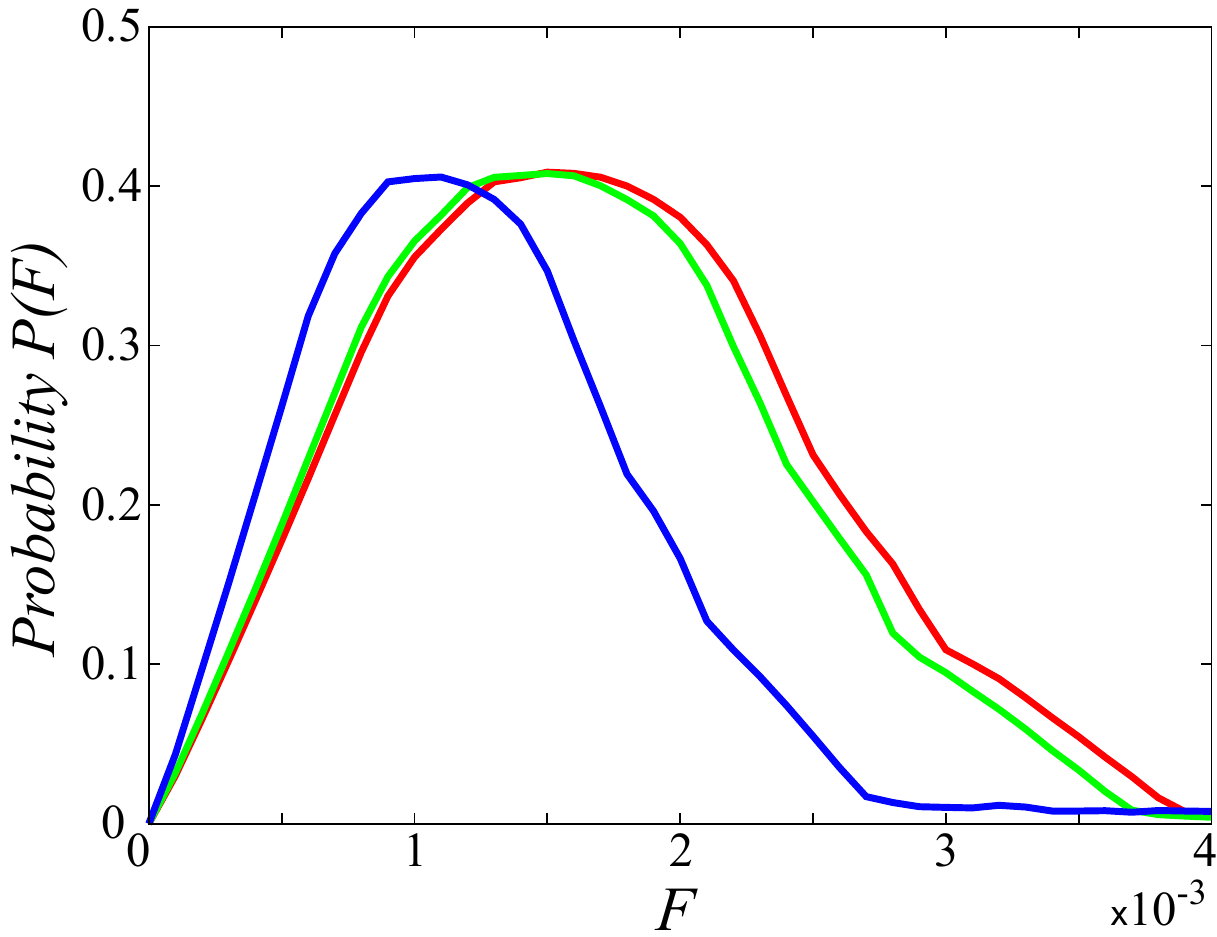}
 \caption{Formation probability as a function of $F$ for an initial energy $E_0=3\times 10^{-9}$ computed from Hamiltonian~\ref{hami}. The parameters of the pulse are $T_{\rm ru}=T_{\rm rd}=5~\mbox{ps}$ and
 $T_{\rm p}=70~\mbox{ps}$ (red line), $T_{\rm ru}=T_{\rm rd}=15~\mbox{ps}$ and
 $T_{\rm p}=70~\mbox{ps}$ (green line) and $T_{\rm ru}=T_{\rm rd}=15~\mbox{ps}$ and
 $T_{\rm p}=140~\mbox{ps}$ (blue line), respectively.}
\label{fig:formation} 
\end{figure}

By the numerical integration of the
equations of motion arising from Hamiltonian \ref{hami}, we propagate the ensemble of trajectories 
for the entire pulse duration.
If after the pulse the energy of a given trajectory is negative, a bound state
is then created. Otherwise, the trajectory remains unbounded.
In our numerical experiments we consider laser pulses with electric field $F$ amplitude
between 0 and  $4\times10^{-3}$ a.u. which corresponds to a laser field of maximal 
intensity of $10^{12}~{\rm W}\cdot{\rm cm}^{-2}$.
The $T_{\rm ru}+T_{\rm p}+T_{\rm rd}$ total duration  of the pulse is taken between 80 ns and 170 ns.
In Fig.~\ref{fig:formation} the formation
probability $P(F)$ as a function of the electric field strength $F$
for three different laser profiles is represented.
Since we start with a positive initial energy, the formation probability is zero for $F=0$. It then
increases sharply with $F$ up to a given critical value of $F$ which depends of the pulse envelope $g(t)$, and
then, it decreases with $F$. Our objective is to analyze the reversal behavior observed in the
formation curves in order to assess the role of the
different parts of the pulse in the building up of this curve.
To this end, we analyze separately the role of the ramp-up, the
plateau and the ramp-down in the dynamics of the system. Special attential is put on
the study of the dynamics during the plateau because this study 
provides important information about the phase space structure of the system and its possible impact in the formation mechanism.
Although results are not being reported here, it is worth noting that, from the computations with ensembles
of trajectories with initial conditions where $P_{\phi}$ and $P_{\theta}$
were not necessarily fixed to zero, the formation
probability has exactly the same shape observed in Fig.3.
In this way, this reversal behavior seems to be very robust and not restrited to
trajectories with initial conditions on the
invariant manifold $P_{\phi}=0$ and with initial conditions $P_{\theta}=0$.
\begin{figure}
 \includegraphics[width=0.5\textwidth]{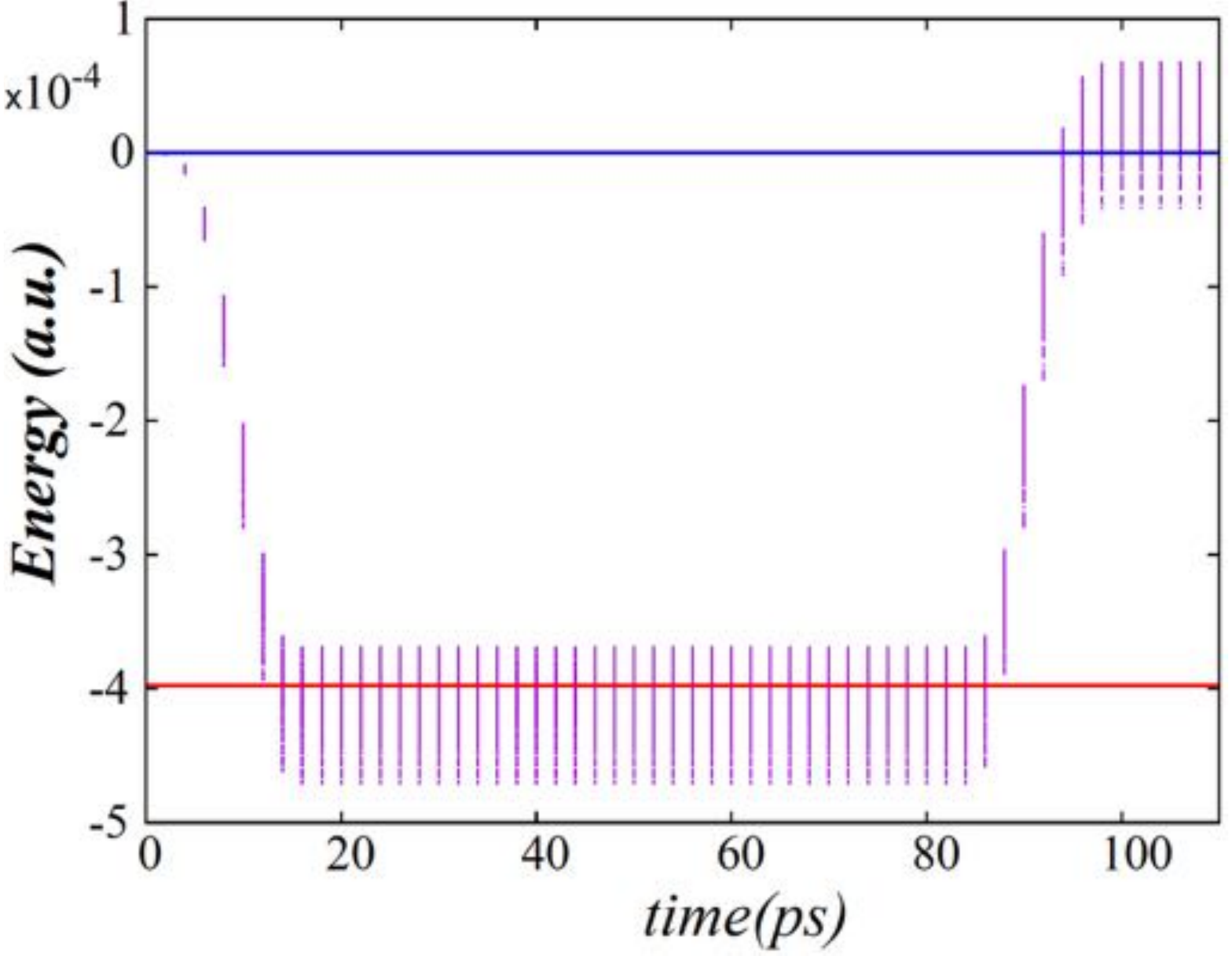}
 \caption{Evolution of the energy of an ensemble of trajectories with initial energy $E_0=3\times 10^{-9}$ a.u.
 The amplitude of the laser field is $F=1.5\times10^{-3}$ a.u.
 The parameters of the pulse are $T_{\rm ru}=T_{\rm rd}=15~\mbox{ps}$ and
 $T_{\rm p}=70$ ps. The red and blue lines indicate the dissociation
 energy \ref{eq:enegydisociacion} and
 the zero energy, respectively.}
\label{fig:energies} 
\end{figure}
\subsection{Role of the ramp-up of the laser pulse}
\begin{figure}
 \includegraphics[width=0.5\textwidth]{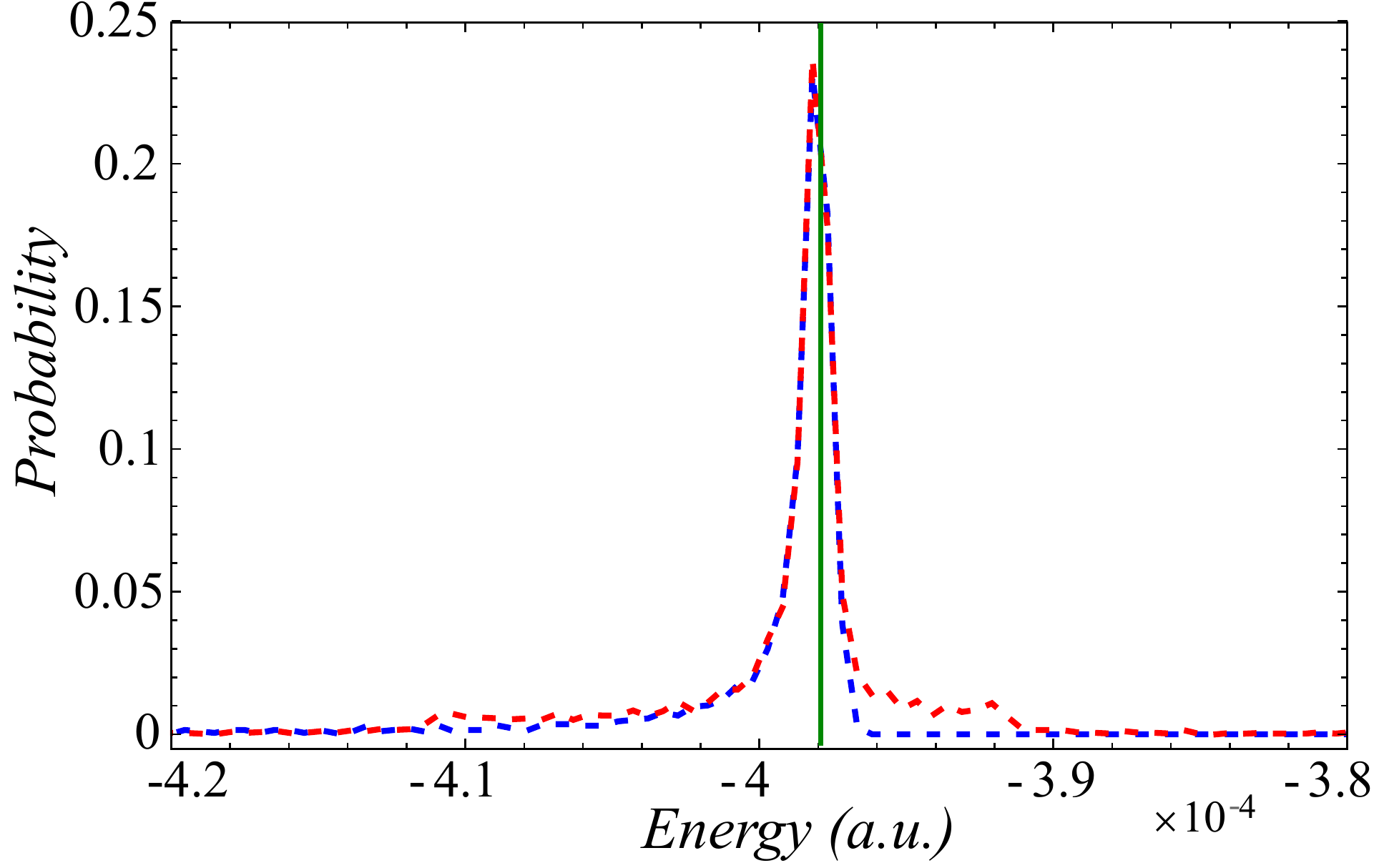}
 \caption{Probability distribution of the energy of an ensemble of trajectories with an initial
 energy $E_0=3\times 10^{-9}$ a.u. after a ramp-up of 15 ps (dashed red line).
 The dashed blue line is the probability energy distribution of an ensemble given by Eq.~\ref{eqn:Ef}.
The dissociation energy
$E_d$ for this electric field is denoted with the
green vertical line.
The parameters of the pulse are $T_{\rm ru}=T_{\rm rd}=15~\mbox{ps}$ and
 $T_{\rm p}=70~\mbox{ps}$ and the amplitude of the electric laser field is $F=1.5\times10^{-3}$ a.u.}
\label{fig:distribution} 
\end{figure}

In Fig.~\ref{fig:energies} the evolution as a function of time of the energy of a
bunch of representative trajectories with initial energy $E_0$ is represented for an
amplitude of the laser field of $F=1.5\times10^{-3}$ a.u.
The parameters of the pulse are $T_{\rm ru}=T_{\rm rd}=15~\mbox{ps}$ and $T_{\rm p}=70$ ns.
As expected, the role of the ramp-up is to decrease the energy of the system and 
to promote the initially unbounded trajectories in a region where, potentially, they might be bounded.
After the ramp-up, the energy probability is represented in Fig.~\ref{fig:distribution} (dashed red line).
\noindent
This energy distribution indicates that, after the ramp-up, a big amount of trajectories acquire energy values
around a relatively narrow region.
This peak structure is easily understood assuming that the dynamics does not play a major role. Under this
assumption, the energy $E_f$ at
the end of the ramp-up is approximately
\begin{equation}
\label{eqn:Ef}
E_f \approx \varepsilon(R)-\frac{F^2}{4}\left[\alpha_\parallel(R) \cos^2\theta+\alpha_\perp(R) \sin^2\theta \right],
\end{equation}
In Fig.~\ref{fig:distribution} the dashed blue line is the
the probability distribution given by $E_f$, where $\theta$ and $R$ are evaluated in the ensemble of
trajectories after the ramp-up. We notice that this distribution displays the same peak structure as
the distribution of energies after the ramp-up computed from the equations of motion
associated with Hamiltonian \ref{hami}. The peak is located around the maximum of
$E_f$ for $R_{\rm max}$,  which is the maximum distance considered in the ensemble of
initial conditions. This maximum of energy almost corresponds to the dissociation energy 
$E_d\approx3.977\times10^{-4}$ a.u. for $F=1.5\times10^{-3}$ a.u. This value is denoted with the
green vertical line in Fig.~\ref{fig:distribution}. Around $R=R_{\rm max}$, the
potential $\varepsilon (R)$ is negligible. This means that
most of the trajectories have energies as if they were at $R=R_{\rm max}$. This comes from the fact that
the potential is rather flat for $R\geq 30$, which affects more than 75\% of the trajectories.
Therefore, the dynamics is very slow for these trajectories, and
$\theta$ and $R$ are approximately constant over the duration of a ramp-up of a few picoseconds.
\begin{figure}
 \includegraphics[width=0.5\textwidth]{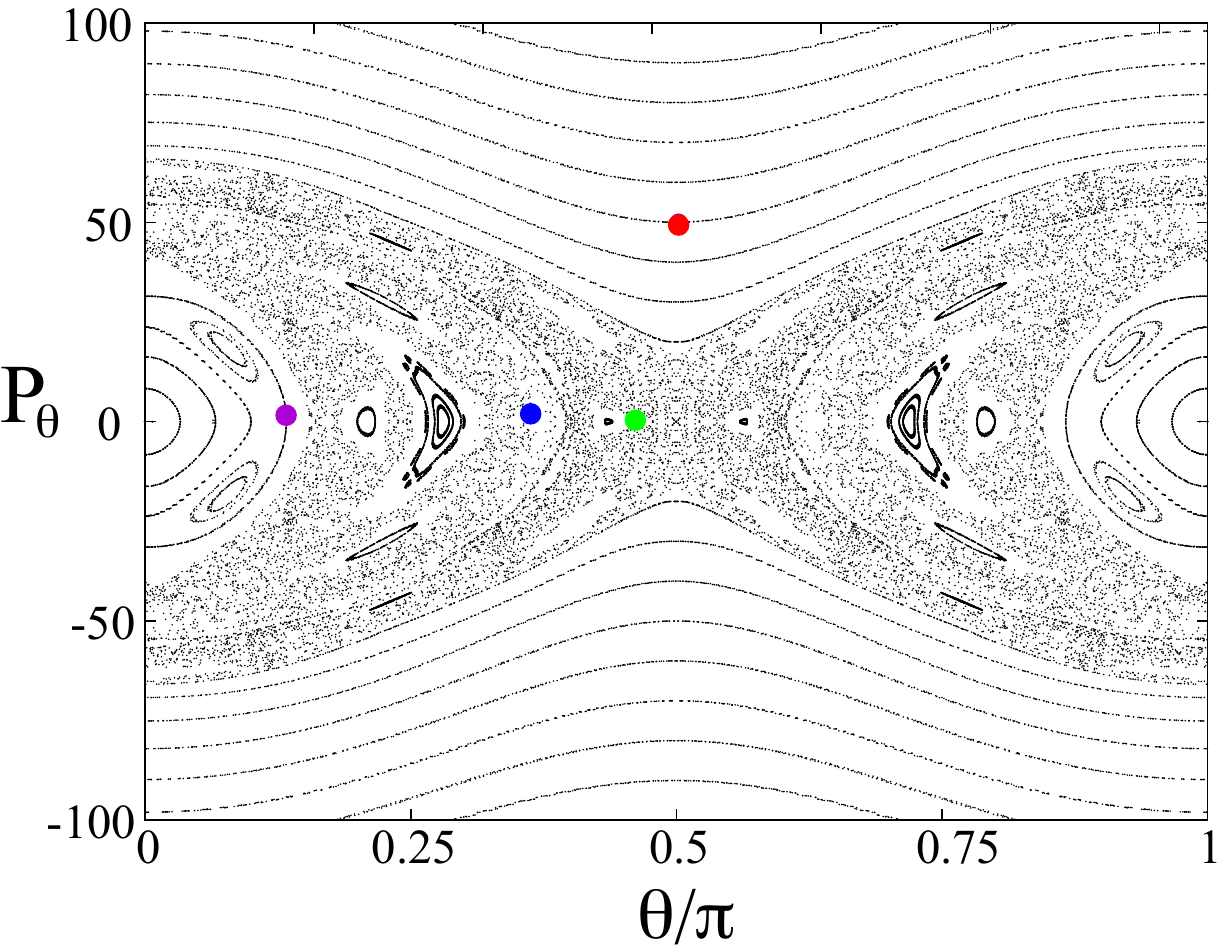}
 \caption{Poincar\'e section ($P_R=0$, $\dot P_R<0$) of Hamiltonian ~\ref{hami}  for
 an energy $E=-3.98\times10^{-4}$ a.u. and for an electric field $F=1.5\times10^{-3}$ a.u.}
\label{fig:sos} 
\end{figure}
 \subsection{Dynamics during the plateau}
During the plateau, Hamiltonian~\ref{hami} is autonomous and with two degrees of freedom. We visualize
the nonlinear dynamics using Poincar\'e surfaces of section.
A convenient Poincar\'e section is $P_R=0$ with $\dot{P_R} >0$, represented in
the plane $(\theta,P_\theta)$.
Since we would like to gain insight into the formation probability, we look at bounded trajectories
for which the distance $R$ is oscillating in time. In addition, to compute the surface of section
we select the value of the most probable energy, i.e., the peak in
Fig.~\ref{fig:distribution} which roughly corresponds to $E=-3.98\times10^{-4}$ a.u.
For a single value of $(\theta,P_\theta)$ there are two possible values
of  $R$, one close to the inner turning point and another one for a larger value of $R$ close
the outer turning point. The first one corresponds to $\dot{P_R}>0$ and the second one to
$\dot{P_R}<0$. In order to draw the Poincar\'e section, we must allow the trajectory
to cross the section a relatively high number of times, so we consider the long term
dynamics, much larger than the duration of the laser pulse. A Poincar\'e section of
Hamiltonian~\ref{hami} is represented on Fig.~\ref{fig:sos}. Each initial condition is
integrated up to $10^5 \mbox{ ps}$.  

\begin{figure}
\includegraphics[width=0.3\textwidth]{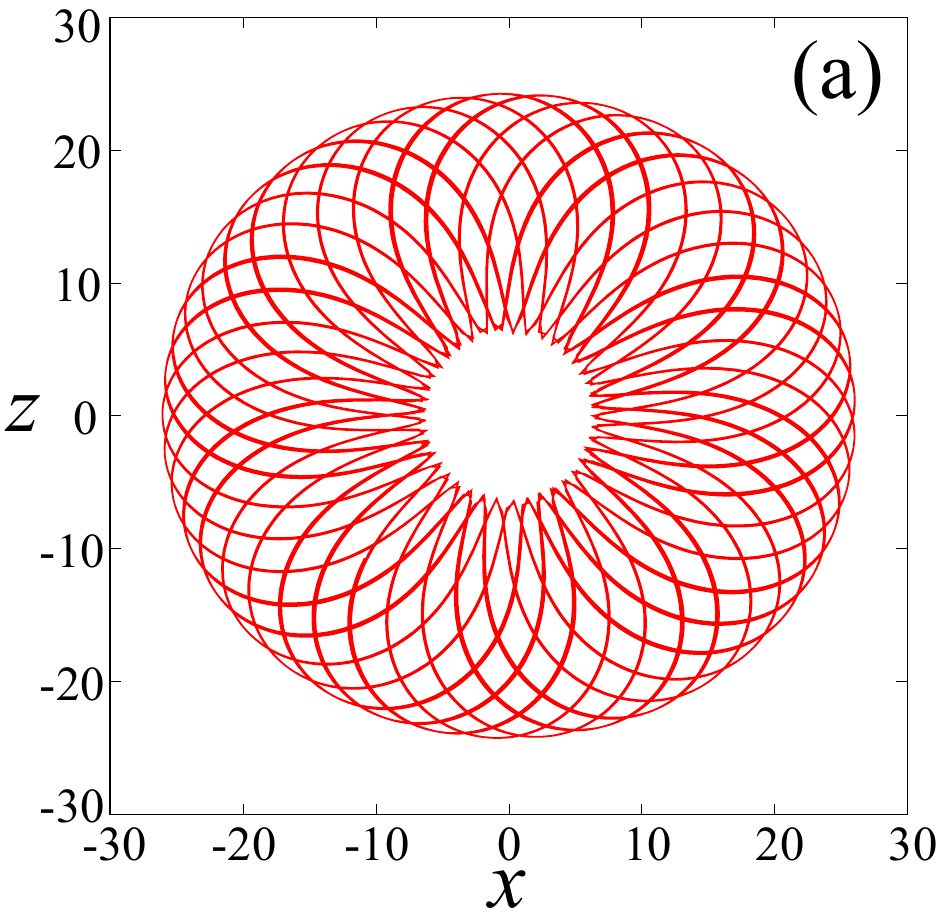} \quad \includegraphics[width=0.3\textwidth]{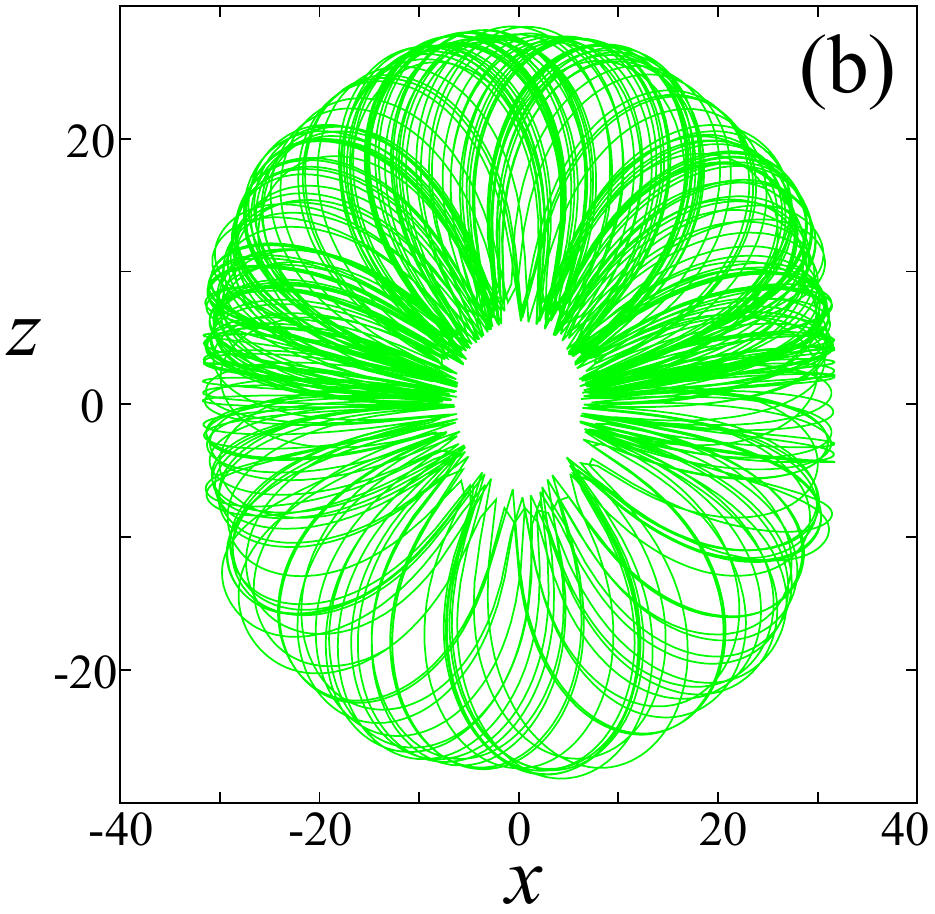}\\[2ex]
 
\includegraphics[width=0.3\textwidth]{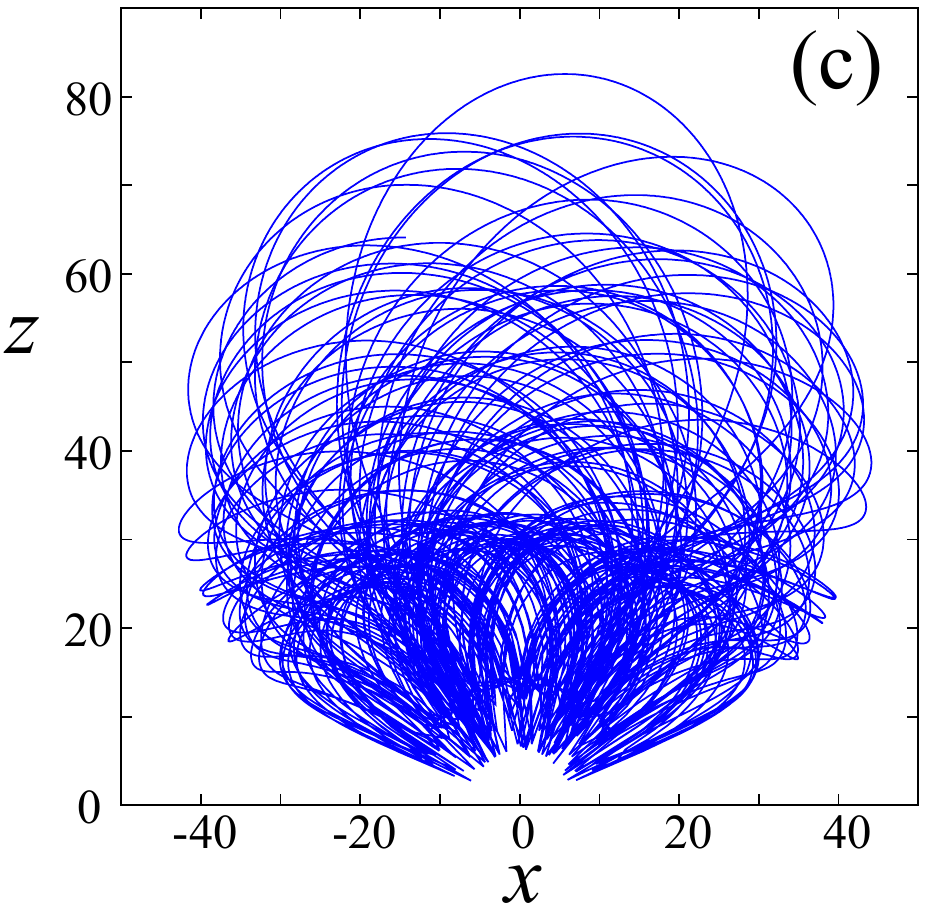} \quad \includegraphics[width=0.3\textwidth]{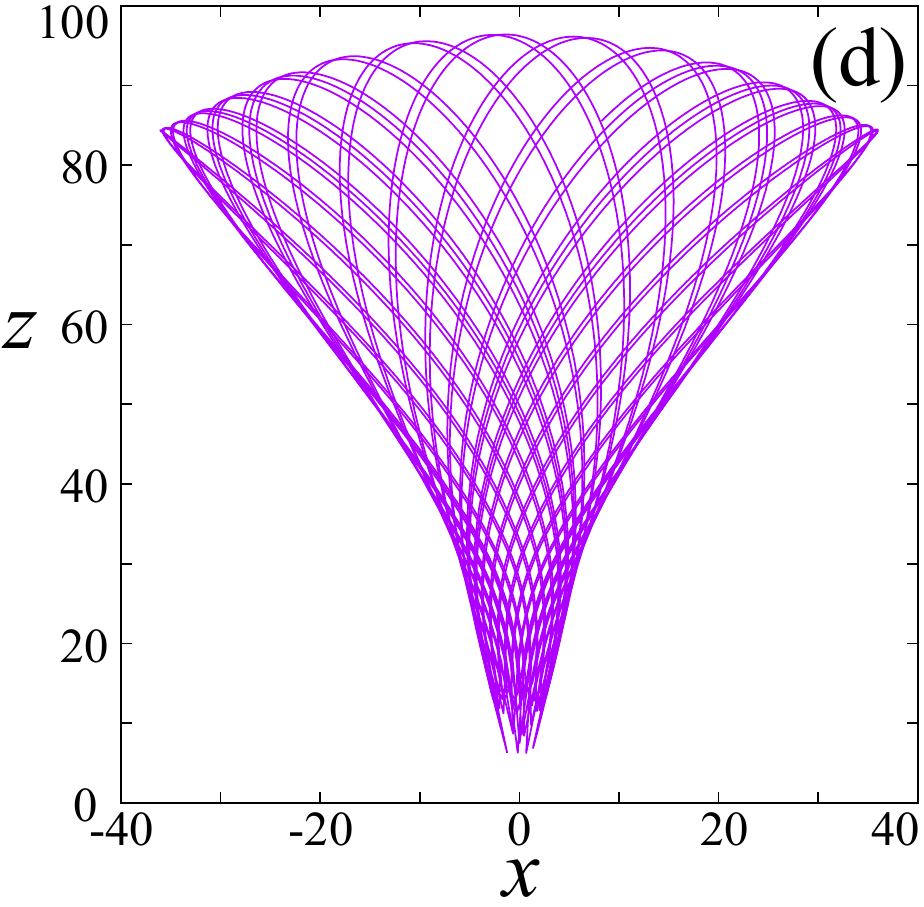}
\caption{Trajectories in the plane $(R\sin\theta,R\cos\theta)$ of Hamiltonian~\ref{hami} for $F=1.5\times 10^{-3}~\mbox{a.u.}$ and energy $E_0=-3.98\times 10^{-4}$ a.u.
(a)  Rotational trajectory with initial conditions $\theta=\pi/2$, $P_\theta=50$ and $P_R=0$ (red dot
in Fig.~\ref{fig:sos});
(b)  rotational chaotic trajectory with initial conditions $\theta=1.45$, $P_\theta=0$ and $P_R=0$
(green dot in Fig.~\ref{fig:sos});
(c)  vibrational chaotic trajectory with initial conditions $\theta=1.1$, $P_\theta=0$ and $P_R=0$
(blue dot in Fig.~\ref{fig:sos})
and (d)  vibrational regular trajectory with initial conditions $\theta=0.2$, $P_\theta=0$ and $P_R=0$
(purple dot in Fig.~\ref{fig:sos}).}
\label{fig:orbits} 
\end{figure}

\noindent
We notice that for a reasonable range of values of $P_\theta$ the dynamics resembles
the one of a forced pendulum with rotational and librational trajectories, and
a ``rotational'' chaotic zone around the hyperbolic point at $\theta=\pi/2$ \cite{PLA2010}. We use the
term ``rotational'' chaotic zone to indicate the chaotic trajectories
spanning the whole interval $[0,\pi]$ for the angle $\theta$. We observe a different ``librational'' chaotic
zone around the elliptic points (located at $\theta=0$ and $\theta=\pi$) which is apparently
disconnected from the rotational chaotic zone, at least on the duration of the numerical integration we have performed.
The elliptic points at $\theta=0, \pi$ correspond to two
straight radial oscillations from $R_{\rm a}$ to $R_{\rm b}$. These values  
$R_{\rm a}$ and $R_{\rm b}$ are the two solutions of $\varepsilon(R)-F^2\alpha_\parallel (R)/4=E$.
We refer to these radial periodic orbits as $I_R$.
In Fig.~\ref{fig:orbits} some sample trajectories are shown. The initial conditions of these
orbits are taken on the surface of section of Fig.~\ref{fig:sos}.
A rotational
trajectory is depicted in Fig.~\ref{fig:orbits}a; these trajectories live on two-dimensional invariant tori.
The orbit in Fig.~\ref{fig:orbits}b is an example of chaotic trajectory in the ``rotational'' chaotic zone.
We notice that the interatomic distances of these two trajectories do not reach large values.
Figure~\ref{fig:orbits}c shows a trajectory in the ``librational'' chaotic zone; indeed, we
notice that the trajectory does not span the whole interval of definition of the angle $\theta$.
Finally, in Figure~\ref{fig:orbits}d a trajectory in a regular elliptic island near
the elliptic fixed point around $\theta=0$ is shown. We notice that these last two trajectories
reach very large values of $R$. 
As expected, all trajectories remain bounded since
the energy $E=-3.98\times 10^{-4}$ a.u. is below the
dissociation energy $E_d\approx3.977\times10^{-4}$ for $F=1.5\times 10^{-3}~\mbox{a.u.}$

\begin{figure}
 \includegraphics[width=0.6\textwidth]{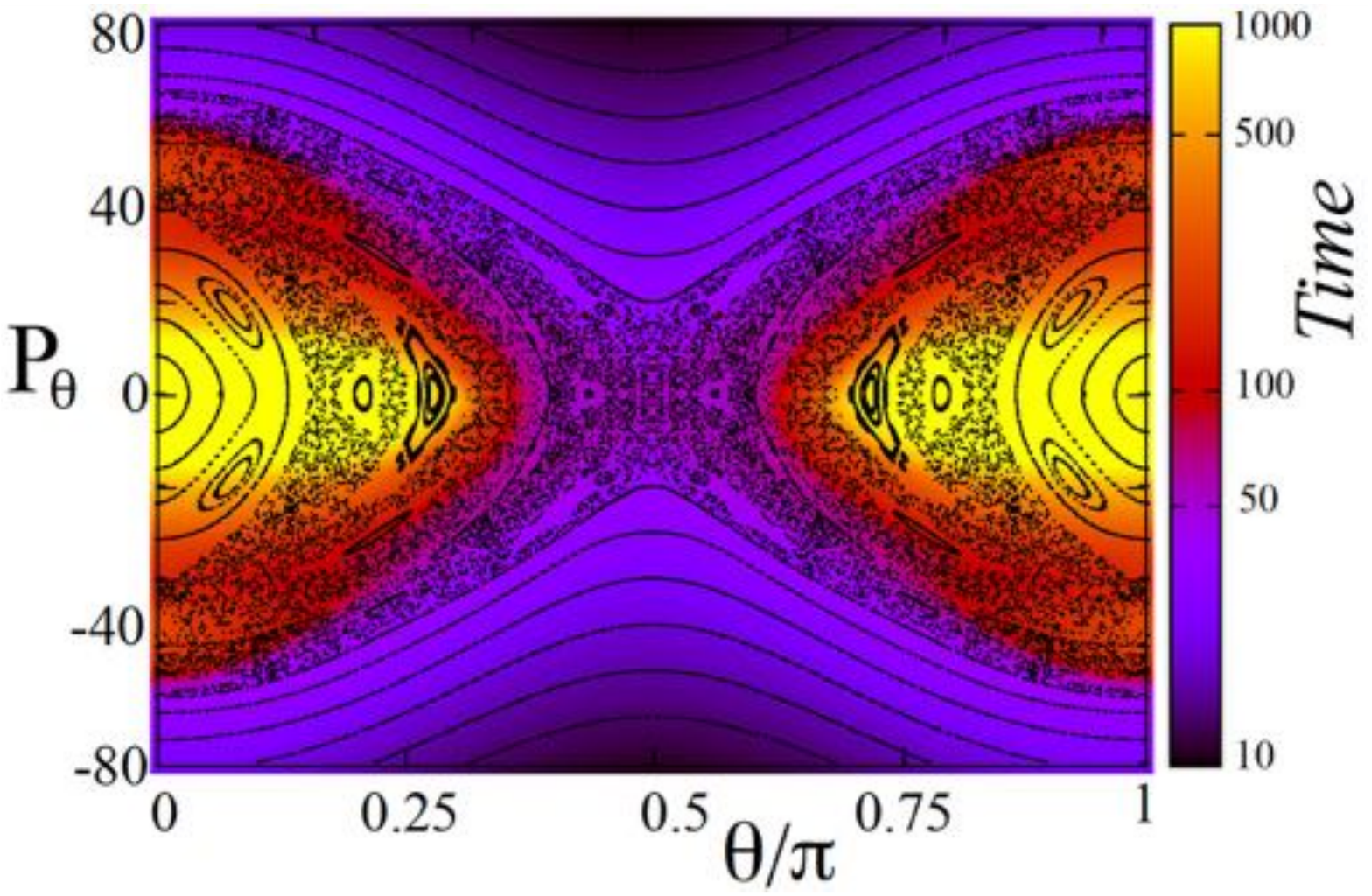}
 \caption{First recurrence time (in ps) in the Poincar\'e section $(P_R=0,\dot{P_R}<0)$ in the plane $(\theta,P_\theta)$ for $F=1.5\times 10^{-3}~\mbox{a.u.}$ and energy $E=-3.98\times 10^{-4}$ a.u.
 The color axis has been saturated at 1000~ps for clarity. In the middle region, the recurrence
time reaches above 1400~ps.
Note that a logarithmic scale is used in the color code.}
\label{fig:return} 
 \end{figure}

\begin{figure}
 \includegraphics[width=0.5\textwidth]{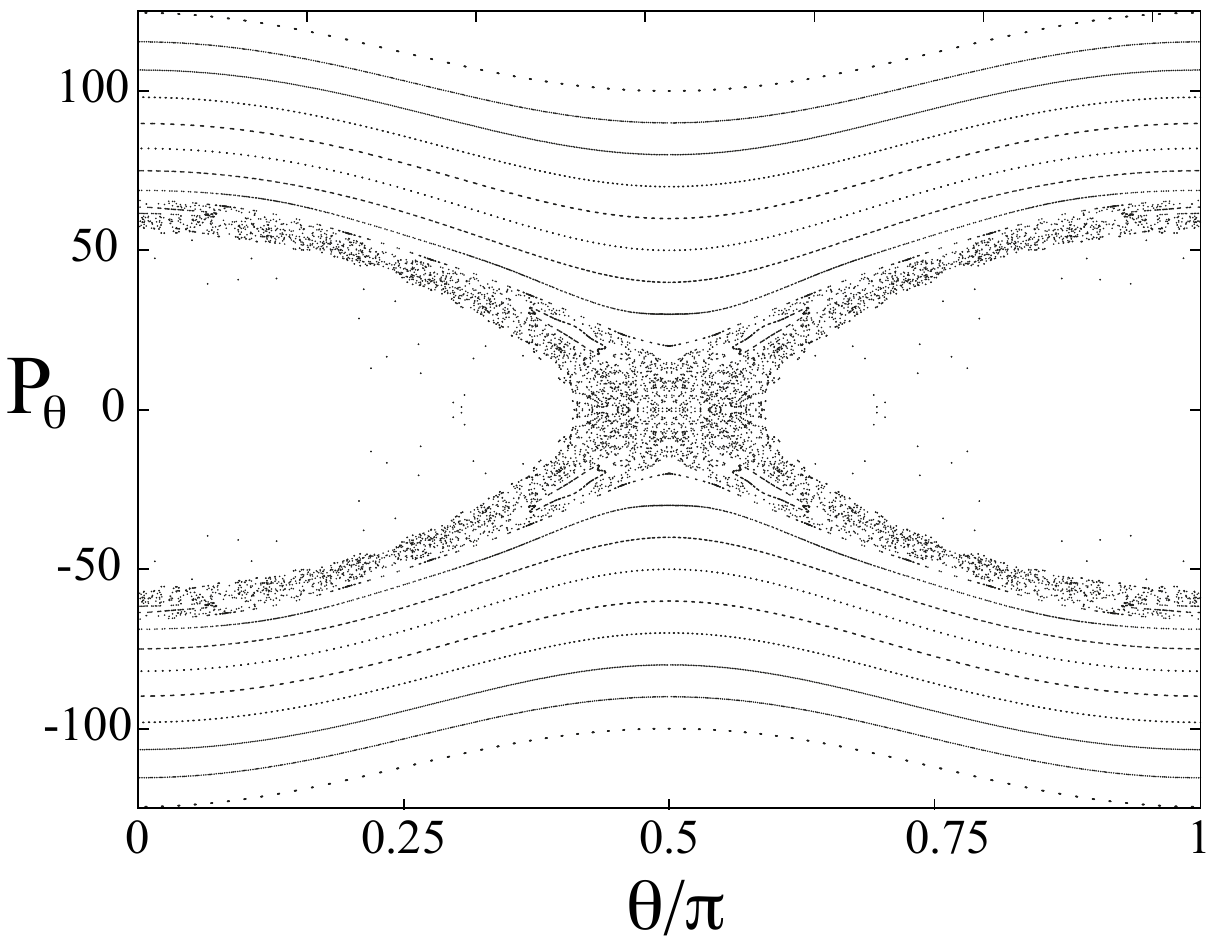}
 \caption{Poincar\'e section ($P_R=0$, $\dot P_R<0$) of Hamiltonian ~\ref{hami}  for
 an energy $E=-3.976\times10^{-4}$ a.u. and for an electric field $F=1.5\times10^{-3}$ a.u.}
\label{fig:sos2} 
\end{figure}

\noindent
What is not apparent in the Poincar\'e section of Fig.~\ref{fig:sos} is the time scales of the
dynamics.  In order to illustrate this property, we
plot the first recurrence time (the time it takes a trajectory to cross the
Poincar\'e section for the first time after starting on the Poincar\'e section) as
a function of $(\theta,P_\theta)$ on the Poincar\'e section. The recurrence time map corresponding to
the surface of section of Fig.~\ref{fig:sos} is shown in Fig.~\ref{fig:return}. 
As we can observe in this color map, in the rotational zones, the dynamics is
rather fast (of the order of tens of ps), while
in the librational zones the dynamics is much slower (on the order of a thousand ps).
This is due to the fact that the trajectories in the librational zones (see Figs.~\ref{fig:orbits}c and
\ref{fig:orbits}d) reach rather large values of $R$ where the potential is extremely flat and hence the
dynamics is potentially extremely slow.
During the plateau of the pulse, for $E<E_d$ [see Eq.~\ref{eq:enegydisociacion}], the trajectories are bounded
and the ones which are the most stretched are around the radial modes $I_R$.
As the energy $E$ gets closer to $E_d$, the maximum
radius $R_{\rm b}$ of $I_R$ increases rapidly. When the energy crosses the 
value $E_d$, the radial trajectories $I_R$ and the quasiperiodic orbits surrounding them are the first orbits
to be unbounded because these orbits are localized along the dissociation channels at $\theta=0, \pi$.
This fact is observed in the Poincar\'e section of Fig.~\ref{fig:sos2} where the holes in the regions
around $\theta=0, \pi$ correspond to the unbounded trajectories.

\subsection{Dynamics during the ramp-down}
\begin{figure}
 \includegraphics[width=0.5\textwidth]{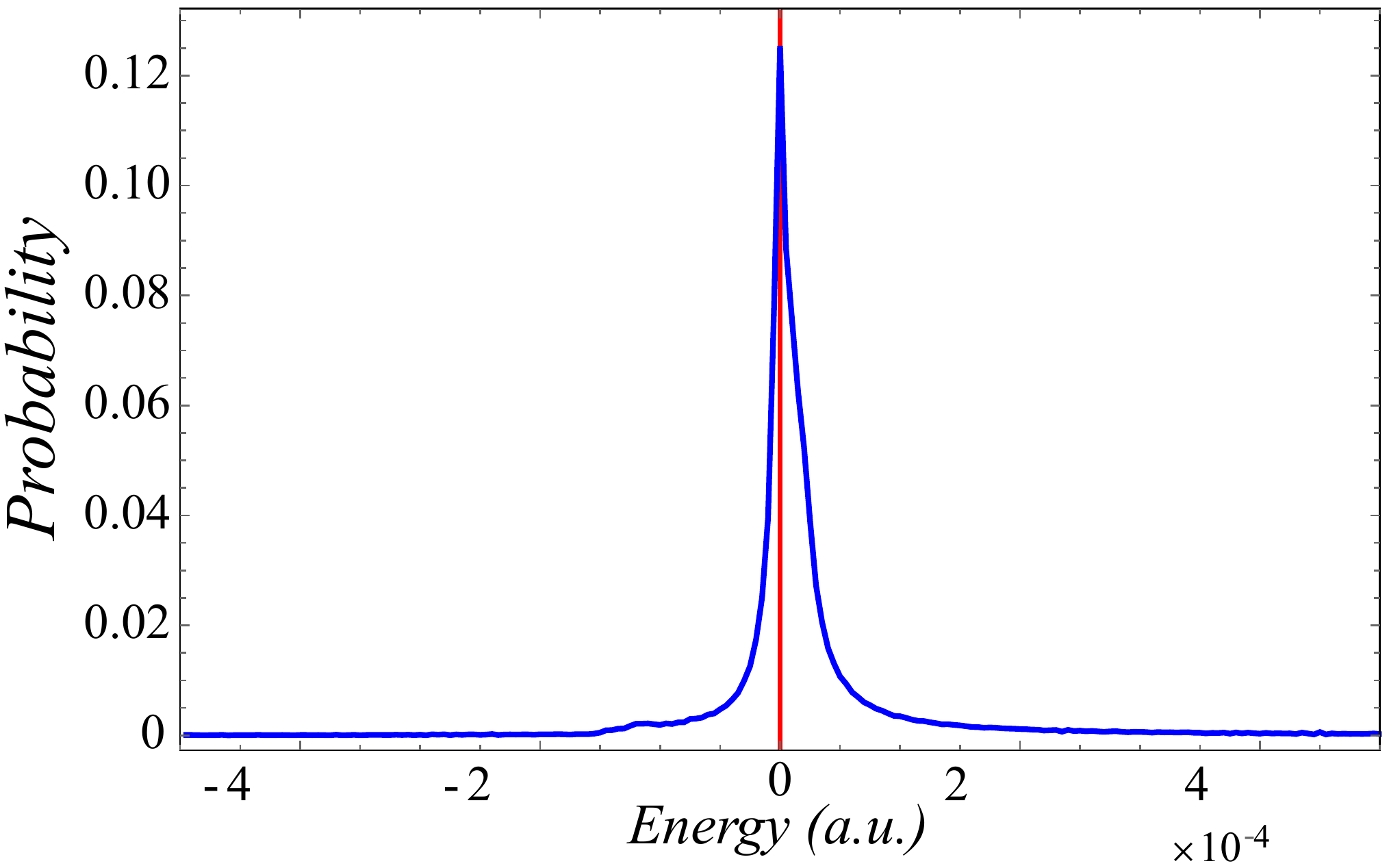}
 \caption{Probability distribution of the energy of an ensemble of trajectories with an initial
 energy $E_0=3\times 10^{-9}$ a.u. after the a ramp-down of 15 ps.
 The vertical red line indicates the zero energy value.
 The amplitude of the laser field is $1.5\times10^{-3}$ a.u.
 and the parameters of the pulse are $T_{\rm ru}=T_{\rm rd}=15~\mbox{ps}$ and
 $T_{\rm p}=70~\mbox{ps}$.}
\label{fig:distribution2} 
\end{figure}

As we observe in Fig.~\ref{fig:energies}, the expected role of the ramp-down is to increase the energy of the 
trajectories. Note that not all the bounded dressed states, i.e., the bounded states in the
presence of the laser field, remain bounded after the ramp-down. When
the energy probability distribution after the ramp-down is calculated (see  Fig.~\ref{fig:distribution2}), we
observe a strong peak structure which indicates that, after the ramp-down, most of the trajectories 
have energies in a narrow region around zero.

\begin{figure}
 \includegraphics[width=0.5\textwidth]{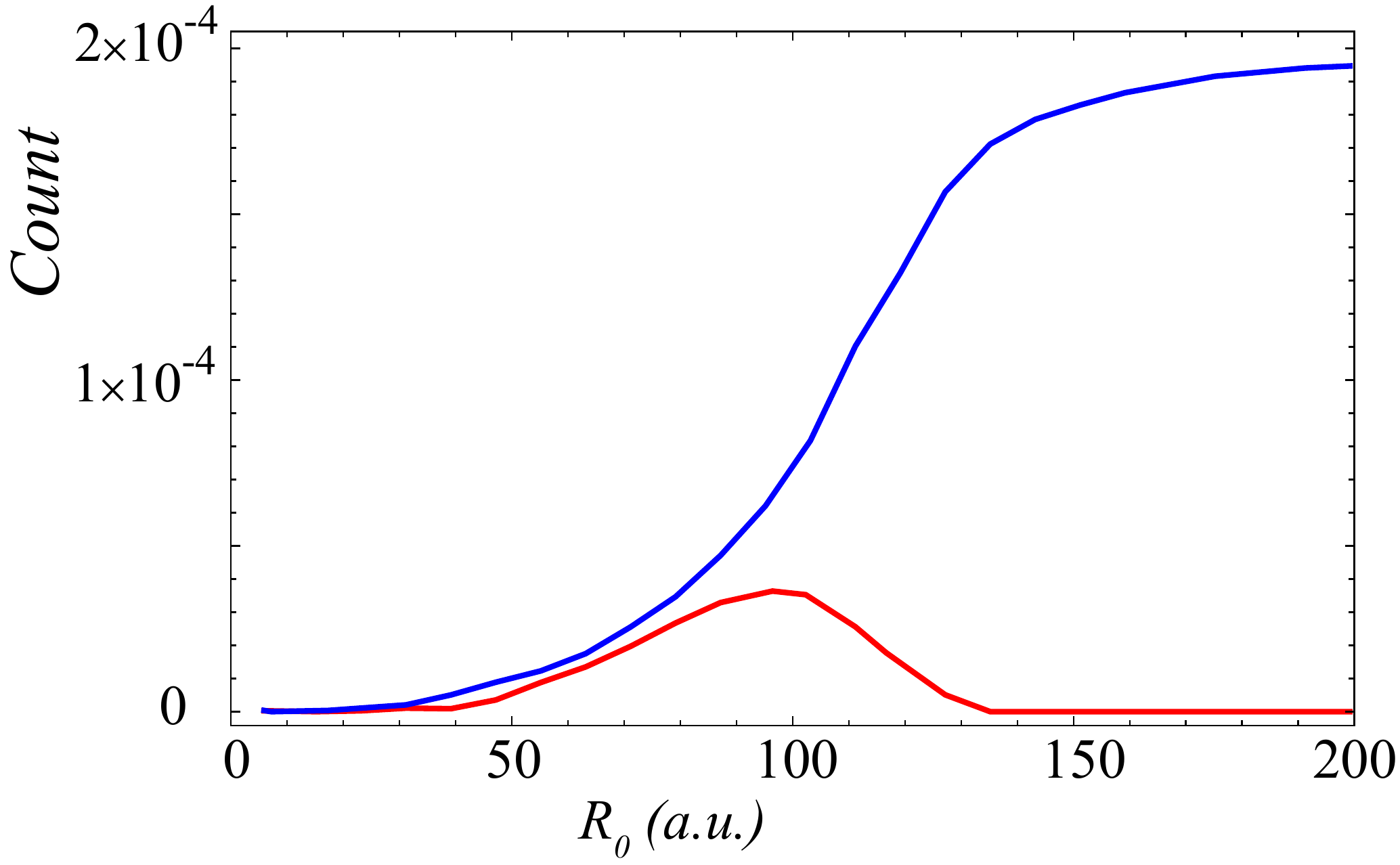}
 \caption{Histogram of the initial conditions leading to formation (red line) and leading to
 dissociation (blue line). The parameters of the laser are  
 $F=1.5\times10^{-3}$ a.u., $T_{\rm ru}=T_{\rm rd}=15~\mbox{ps}$ and
 $T_{\rm p}=70~\mbox{ps}$.
 The energy of the trajectories is $E_0=3\times 10^{-9}$ a.u. }
\label{fig:histogramaTrappedDisso} 
\end{figure}

Where are the formed trajectories in phase space? This is a particularly difficult question
to address since, besides the dependence of the formed trajectories with the initial conditions, it
highly depends on the parameters of the laser pulse (like the intensity, the duration of the ramp-up,
plateau and ramp-down). In particular, it is not possible to predict on the
Poincar\'e section represented in Fig.~\ref{fig:sos} which initial conditions lead to formation and which
ones to dissociation. The main reason is that, depending on the duration of the pulse, the same initial condition can lead to formation or dissociation. One of the noticeable features is that the formed trajectories
have a finite range for the distance, meaning that if the distance between the two atoms is
too large, it will not lead to formation. For instance, for $F=1.5\times 10^{-3}$, this maximum distance is about
130~a.u.\ In Fig.~\ref{fig:histogramaTrappedDisso} we represent the histogram of initial
distances leading to formation compared to the one which lead to dissociation, where we notice
that after some fixed initial distance, the formation is no longer possible. 
We also notice that the trajectories leading
to formation are the ones with small values of $P_\theta$, especially at the
end of the laser pulse.
From the pendulum-like structure of the Poincar\'e map of Fig.~\ref{fig:sos}, we know that
the phase space is populated with two main types of trajectories, namely, vibrational and rotational
trajectories. As we illustrate in Fig.~\ref{fig:orbits}, the vibrational orbits reach the largest interatomic distances. Thence, because the dimer must be formed with trajectories connecting
large and small values of $R$ and most of the orbits have initial conditions with
values of $R_0>25$ a.u., we can argue
that vibrational trajectories should play a dominat role in the formation mechanism.
Moreover, because the radial mode $I_R$ is the simplest vibrational orbit, it is expected to find in
this periodic orbit the same qualitative formation behavior observed in the full system.
In other words, this information allows one to 
focus on the formation dynamics arising from the one degree of freedom Hamiltonian
associated with $I_R$, e.g., with  a Hamiltonian model where the
degree of freedom $(\theta,P_\theta)$ is frozen.

\section{One degree of freedom model}
The co-dimension 2 manifolds defined by $\theta=k \pi/2$ ($k=0,1,2$) and
$P_\theta=0$ are invariant under the dynamics. This allows us to define essentially
two reduced Hamiltonian systems with 1+1/2 degrees of freedom:   
\begin{equation}
\label{eq:Ham2D_1}
H_1(R,P_R,t)=\frac{P_R^2}{2\mu}+\varepsilon(R)-g(t)\frac{F^2}{4}\alpha_\parallel(R), \quad 
\mbox{for} \ \theta=0, \pi
\end{equation}
and 
\begin{equation}
\label{eq:Ham2D_2}
H_2(R,P_R,t)=\frac{P_R^2}{2\mu}+\varepsilon(R)-g(t)\frac{F^2}{4}\alpha_\perp(R), \quad 
\mbox{for} \ \theta=\pi/2.
\end{equation}
The model~\ref{eq:Ham2D_1} describes the dynamics of the radial mode $I_R$ and it is
structurally stable, in the sense
that if we move slightly away from this model by considering the full model in a
range of values of $\theta$ and $P_\theta$ close to zero, the dynamics stays in the vicinity of
the ones obtained with the model \ref{eq:Ham2D_1}. On the contrary, the second model described by Hamiltonian~\ref{eq:Ham2D_2} is structurally unstable since trajectories nearby
$\theta=\pi/2$ and $P_\theta=0$ tend to move away from these values in the
full model.  In this way, in what follows we
focus on Hamiltonian \ref{eq:Ham2D_1}.
The corresponding equations of motion are:
\begin{eqnarray}
\label{ecumovi1D}
\dot R &=& \frac{P_R}{\mu},
\nonumber \\
& &\\
\dot P_R & = & -\frac{d \varepsilon(R)}{d R} + g(t)  \frac{F^2}{4} \frac{d \alpha_\parallel(R)}{d R}.\nonumber
\end{eqnarray}
We consider an ensemble of initial conditions $(R_0, P_R^0)$
with energy $E_0=3\times 10^{-9}$ a.u. defined as
$$
E_0=\frac{P_{R_0}^2}{2\mu}+\varepsilon(R_0),
$$
where the initial values of intermolecular distance $R_0$ are distributed in the interval
$[R_{\rm min},R_{\rm max}]=[6.2319, 100]$ a.u. according to the criterion
described in Sec.~\ref{driving}.
\begin{figure}
 \includegraphics[width=0.4\textwidth]{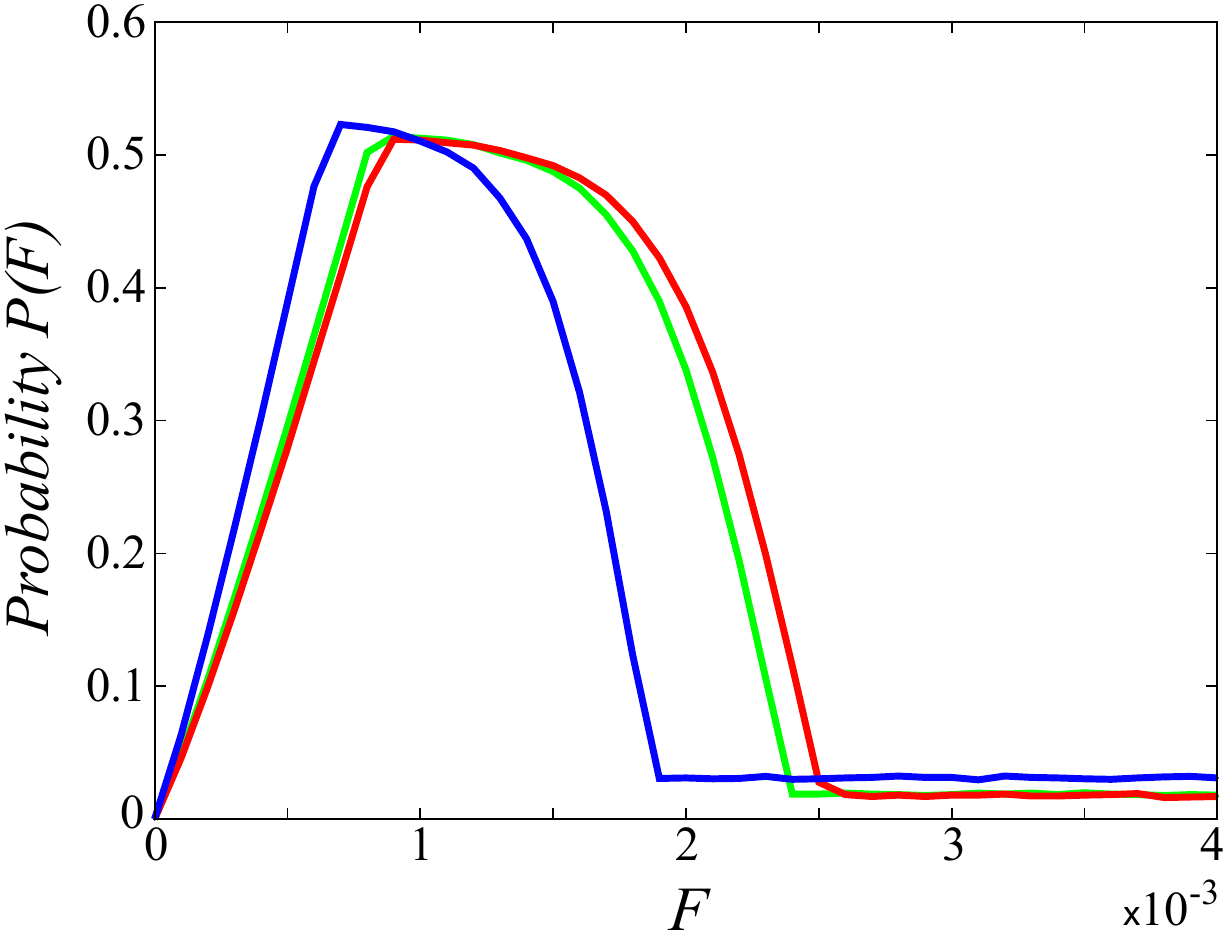}
 \caption{Formation probability as a function of $F$ for an initial energy $E_0=3\times 10^{-9}$ a.u.
 computed using Hamiltonian~\ref{eq:Ham2D_1}. The parameters of the
 pulse are $T_{\rm ru}=T_{\rm rd}=5~\mbox{ps}$ and
 $T_{\rm p}=70~\mbox{ps}$ (red line), $T_{\rm ru}=T_{\rm rd}=15~\mbox{ps}$ and
 $T_{\rm p}=70~\mbox{ps}$ (green line) and $T_{\rm ru}=T_{\rm rd}=15~\mbox{ps}$ and
 $T_{\rm p}=140~\mbox{ps}$ (blue line), respectively.}
 \label{fig:formation2D} 
 \end{figure}
Using this ensemble of initial conditions, we compute the formation probability as a function of the
electric field parameter $F$ and the results are shown in Fig.~\ref{fig:formation2D}.
We notice that we find the same qualitative behavior as in the formation probability for the full Hamiltonian~\ref{hami}, notably the decrease of the probability for sufficiently large amplitudes.

\begin{figure}
 \includegraphics[width=0.4\textwidth]{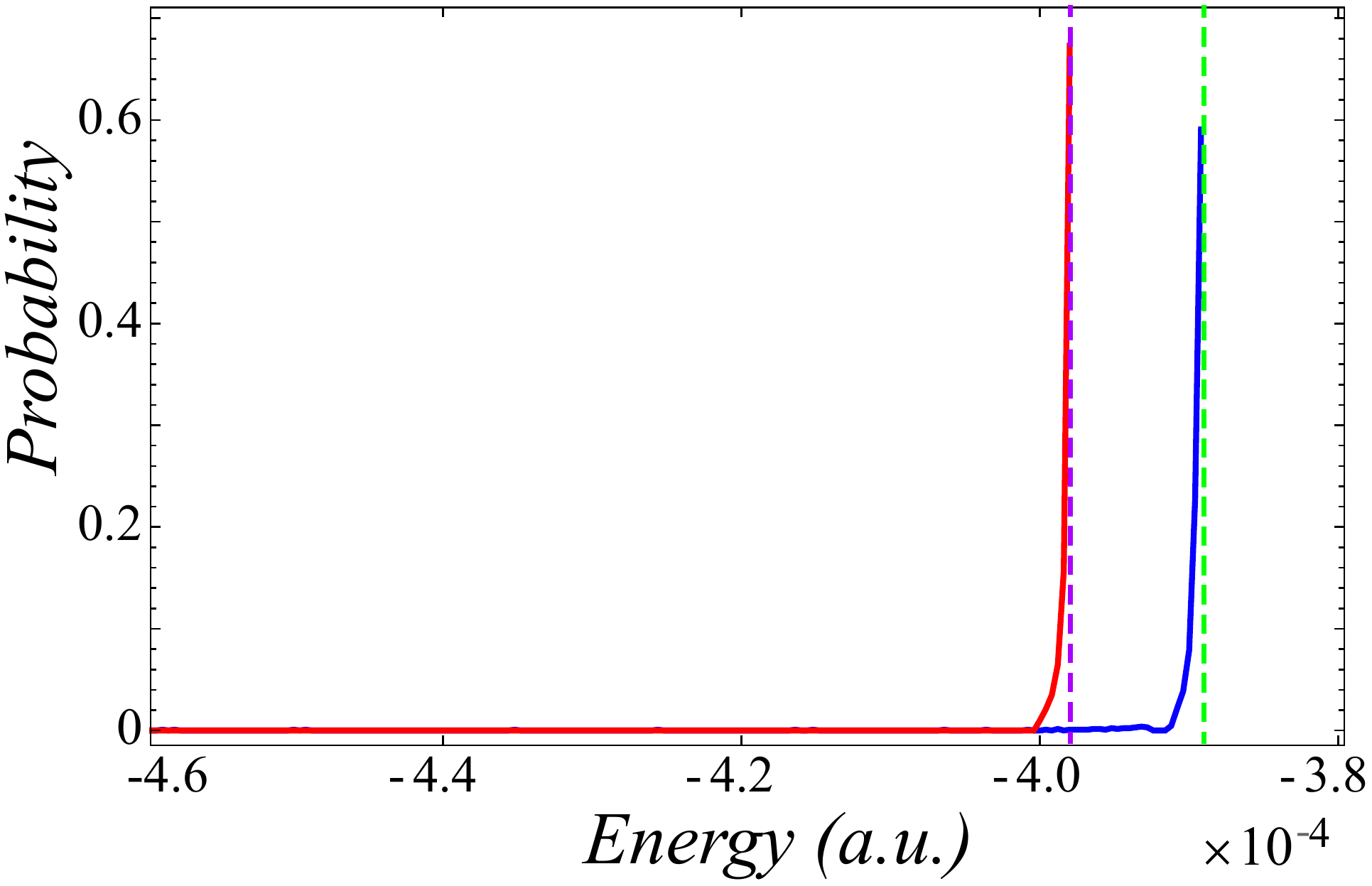}
 \caption{Probability distributions of an ensemble of trajectories with an initial
 energy $E_0=3\times 10^{-9}$ a.u. after a ramp-up of $15~\mbox{ps}$ obtained
 with formula \ref{eq:deltaE}) (red line)
 and with  Hamiltonian~\ref{eq:Ham2D_1} (blue line).
 The amplitude of the laser field is $F=1.5\times 10^{-3}~\mbox{a.u.}$ The dashed green
line indicates the dissociation energy $E_d=-F^2\alpha_\parallel{(\infty)}/4$ while the
dashed purple line denotes the energy $E_d=-F^2\alpha_\parallel{(R_{\rm max})}/4$.}
 \label{fig:distribution4} 
 \end{figure}

\medskip
After a ramp-up of $15~\mbox{ps}$, the probability distribution of the energy is represented in
Fig.~\ref{fig:distribution4} for the value $F=1.5\times 10^{-3}$ a.u. for which
a significant formation probability is observed (see Fig.~\ref{fig:formation2D}).
From the computation of the probability distribution of the energy after the ramp-up
(red line in Fig.~\ref{fig:distribution4}), we observe
again a strong peak structure which indicates that, after the ramp-down, most of the trajectories 
have energies in a narrow region below the dissociation threshold
$E_d=-F^2\alpha_\parallel{(\infty)}/4$.
This is an expected behavior since the effect of the
ramp-up is to decrease the initial energy $E_0$ of the trajectories and due to fact that
$E_0$ is small,  the energies of the trajectories after the ramp-up are
below $E_d$.
 
\medskip
Since the initial distances $R_0$ of our trajectories are in general large, we
assume that, during the ramp-up, the intermolecular distances $R$ do not change
significantly since $\dot{R}=P_R/\mu$ is small. Under this assumption, an
approximation of the momentum at the end of the ramp-up is obtained by considering that $R$ is constant.
Indeed, using the equations of motion \ref{ecumovi1D}, the variation of the radial
momentum induced by the ramp-up of the field is approximately given by
\begin{equation}
\Delta P_R(T_{\rm ru}) \approx \frac{F^2}{8}T_{\rm ru} \frac{d \alpha_\parallel(R_0)}{d R},
\label{deltaP}
\end{equation}
\noindent
where the term of order $F^4$ is neglected and we assume that $d \varepsilon(R_0)/d R \approx 0$.
Since $d \alpha_\parallel(R_0)/d R$ is negative (see Fig.~\ref{fi:curvas}) for most of the values of
$R_0$, we conclude that, in general,
the momentum decreases as a result of the ramp-up. In order to have an approximate
value of the energy at the end of the ramp-up of the laser field for large values of $R_0$, we insert
Eq.~\ref{deltaP} into Hamiltonian \ref{eq:Ham2D_1}.
After neglecting the term of order $F^4$, we get
\begin{equation}
E_{\rm ru}\approx E_0-\frac{F^2}{4}\alpha_\parallel (R_0)+\frac{F^2 T_{\rm ru}}{8 \mu} P_R^0\alpha_\parallel'(R_0). 
\label{eq:deltaE}
\end{equation}
In order to check the validity of the above equation, we compute the probability
distribution of the energy for our set of initial conditions by using Eq.\ref{eq:deltaE}.
The result (blue line in Fig.~\ref{fig:distribution4}) is rather accurate since the probability distribution
obtained from Eq.~\ref{eq:deltaE} is closely peaked below the value
$E=-F^2\alpha_\parallel{(R_{\rm max})}/4$.

\begin{figure}
 \includegraphics[width=0.4\textwidth]{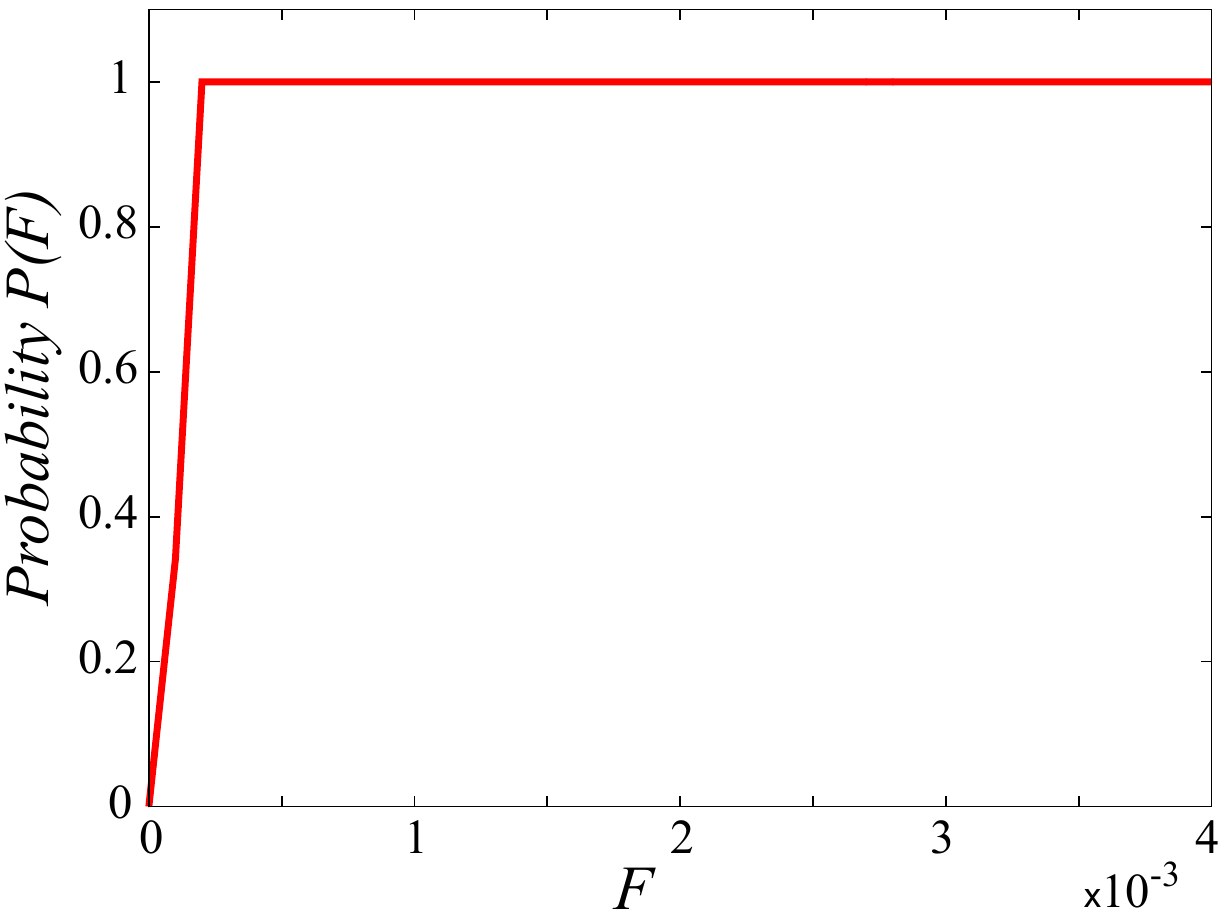}
 \caption{Formation probability for Hamiltonian~\ref{eq:Ham2D_1}
 as a function of $F$ for an initial energy $E_0=3\times 10^{-9}$ a.u. The parameters of
 the pulse are $T_{\rm ru}=15~\mbox{ps}$, $T_{\rm p}=70~\mbox{ps}$ and no ramp-down.}
 \label{fig:proba2Dplat} 
 \end{figure}

\medskip
During the plateau, the Hamiltonian~\ref{eq:Ham2D_1} has one degree of freedom and
the energy of the system is conserved. Since for 
relevant values of $F$, all the energies are below the dissociation threshold
$E_d=-F^2\alpha_\parallel{(\infty)}/4$, all the trajectories remain bounded during the plateau. 
This is confirmed in Fig.~\ref{fig:proba2Dplat} where the formation probability, computed from an energy criterion $E<E_d=-F^2\alpha_\parallel{(\infty)}/4$ is represented as a function of $F$.
\noindent
It means that at all times, all the dimers remain bounded in the presence of the laser field for $F\gtrsim 2\times 10^{-4}\mbox{ a.u.}$, whether a distance or an energy criterion is used.
During the plateau, all the bounded trajectories are periodic and their periods are given by
\begin{equation}
\label{periodo}
T(E,F)=\sqrt{2\mu} \int_{R_{\rm a}(E,F)}^{R_{\rm b}(E,F)}\frac{{\rm d} R}{\sqrt{E-\varepsilon (R)+\frac{F^2}{4}\alpha_\parallel(R)}},
\end{equation}
\noindent
where $R_{\rm a}<R_{\rm b}$ are the two turning points given by the solutions of 
\[
\varepsilon (R) -\frac{F^2}{4}\alpha_\parallel(R)=E<E_d.
\]
\begin{figure}
 \includegraphics[width=0.5\textwidth]{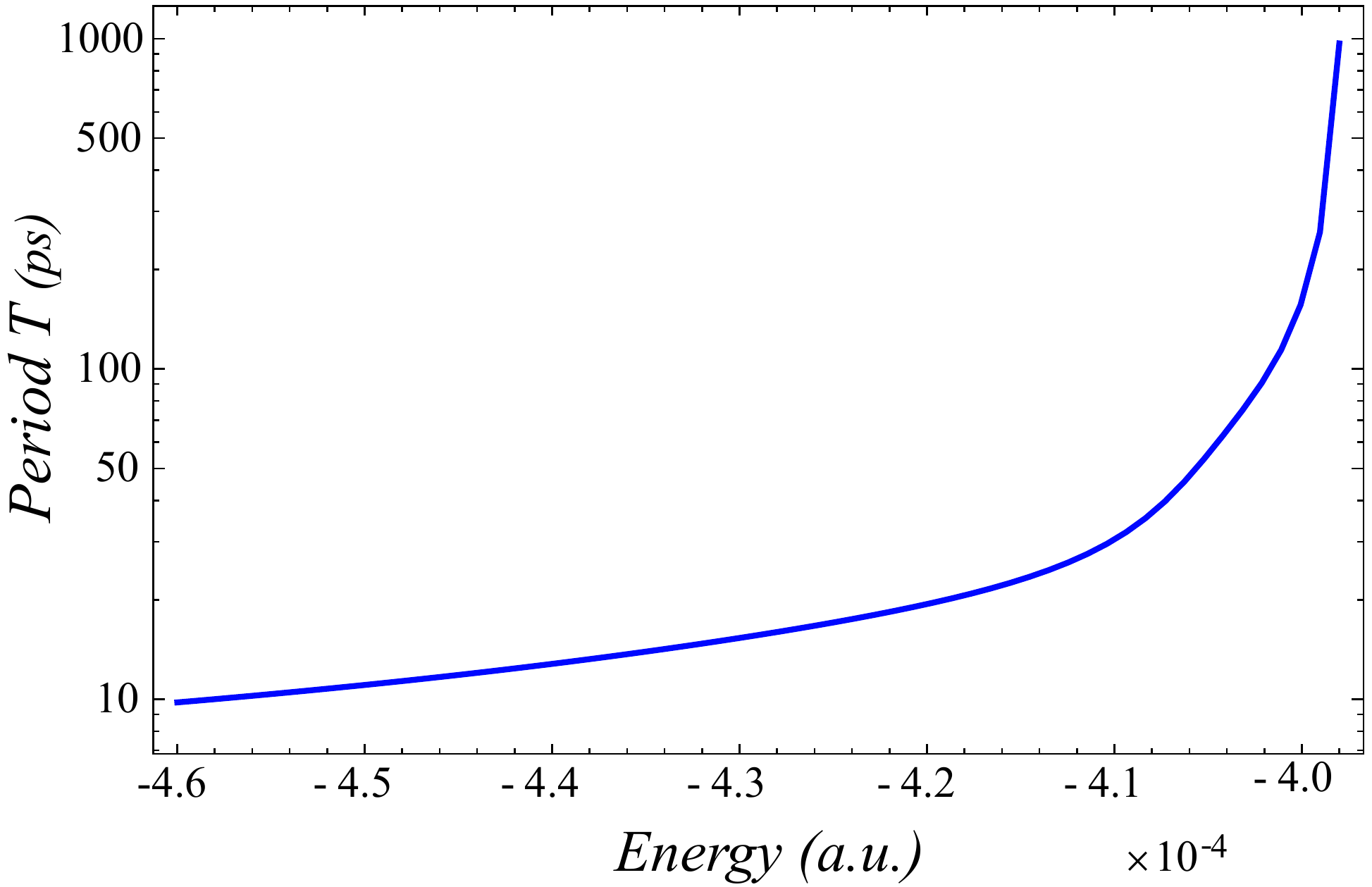}
 \caption{Periods of our ensemble of trajectories for $F=1.5\times10^{-3}$ a.u. using Eq.~\ref{periodo}.
 Note the logarithmic scale in the vertical axis.}
 \label{fig:periods} 
\end{figure}
\noindent
Since the ramp-up promotes most of the trajectories very close but below the threshold
energy values $E_d$, we have computed the periods of our
ensemble of trajectories for $F=1.5\times10^{-3}$ a.u. The
results are shown in Fig.~\ref{fig:periods}. As expected, the motion is very slow in
comparison with the duration of the pulse and it mirrors the observation
made in the first recurrence time map of Fig.~\ref{fig:return}.

\medskip
As we have observed, after the ramp-up and for relevant values of $F$, most of the trajectories remain
bounded during the plateau. However not all these bounded dressed states, i.e., the bounded
states in the presence of the laser field, remain bounded after the ramp-down. 
Even for this one dimensional model it is cumbersome to untangle the
effects of the various parts of the pulse
and to provide insights into the role of the parameters of the pulse.
In order to unravel the dynamics, we consider
 the long-range dynamics of the one degree of freedom Hamiltonian model \ref{eq:Ham2D_1}.

 \section{Simplified potential}
In order to investigate the long-range behavior of Hamiltonian \ref{eq:Ham2D_1}, we assume
that, for $R$ large, the
expressions of the functions defining the potential are [see Eq.~\ref{pecLR} and Eq.~\ref{polaParaLR}],
\begin{eqnarray}
\label{pec}
\varepsilon(R) & \approx & -\frac{b_6}{R^6},
\\[2ex]
\label{para2}
\alpha_{\parallel}(R) &\approx & \alpha_{\rm RbCs} +
\frac{d_2}{R^2} +  \frac{d_3}{R^3},
\end{eqnarray}
and the simplified long-range Hamiltonian becomes
\begin{equation}
\label{eq:Ham_s}
 H_{\rm s}=\frac{P_R^2}{2\mu} -\frac{b_6}{R^6} - g(t) \frac{F^2}{4} (\alpha_{\rm RbCs} +
\frac{d_2}{R^2} +  \frac{d_3}{R^3}).
\end{equation}

\begin{figure}
 \includegraphics[width=0.4\textwidth]{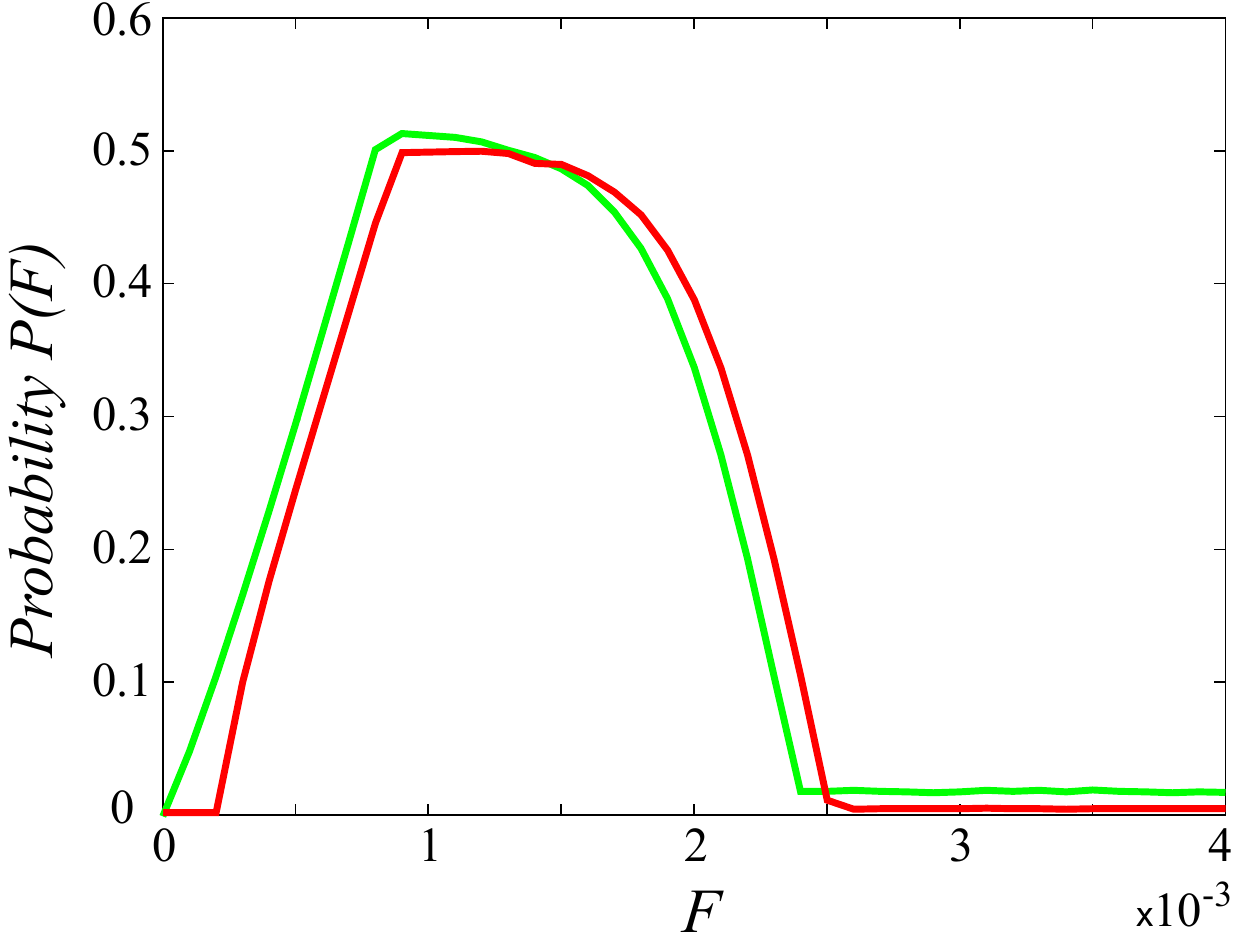}
 \caption{\label{fig:proba1dLR} Formation probability as a function of $F$ for an initial
 energy $E_0=3\times 10^{-9}$ a.u. obtained from the long-range Hamiltonian~\ref{eq:Ham_s} (red line) 
 and the full Hamiltonian \ref{eq:Ham2D_1} (green line).
 The parameters of the pulse are $T_{\rm ru}=T_{\rm rd}=15~\mbox{ps}$ and
 $T_{\rm p}=70~\mbox{ps}$.}
 \end{figure}

The formation probability computed using Hamiltonian~\ref{eq:Ham_s} as a function of $F$ is shown 
in Fig.~\ref{fig:proba1dLR}. This formation probability (green line in Fig.~\ref{fig:proba1dLR}) is in very
 close agreement with
the formation probability obtained with the full Hamiltonian~\ref{eq:Ham2D_1}
(red line in Fig.~\ref{fig:proba1dLR}),
which validates the approximate expressions \ref{pec}-\ref{para2} of the potentials. 
\medskip\noindent
In order to get some insight into this probability curve, we compute the momentum transfer
during the laser pulse as:
\begin{equation}
\label{deltaPP}
\Delta P_R= \frac{F^2}{4} \int\limits_0^{T_{\rm ru}+T_{\rm p}+T_{\rm rd}} g(t) \ \frac{d \alpha_\parallel(R)}{d R} \ dt.
\end{equation}
where we again assume that $d \varepsilon(R)/d R \approx 0$.
Initially, the momentum is given by
$$
P_R^0=\pm \sqrt{2\mu \left[E_0-\varepsilon(R_0) \right]}.
$$
For example, for $R=50$ a.u. the initial value of the momentum is $P_R^0\approx 0.3$ a.u. and
the radial velocity
is  $\dot{R}(0) \approx 3\times 10^{-6}$ a.u. As a consequence, $\dot{R}(0)$ is small
and, therefore, it is reasonable (at least at the leading order) to assume that $R$ is
approximately constant. Using this assumption, the shape of the laser pulse given by
Eq.~\ref{eq:pulse} and the expression \ref{deltaPP}, the momentum transfer induced by the pulse is given by
\begin{equation}
\label{momentumTransfer}
\Delta P_R = \frac{F^2 (T_{\rm ru} + 2 T_{\rm p} + T_{\rm rd})}{8} \ \frac{d \alpha_\parallel(R)}{d R}. 
\end{equation}
We notice that $\Delta P_R < 0$ since $d \alpha_\parallel(R)/d R$ is always negative.
This small momentum transfer, which is of the same order as $P_R^0$, is responsible for the
formation, even though this momentum transfer does not have significantly impact on the
variation of the interatomic distance on the short-time scale of the laser pulse. 
Furthermore, the dependence of the momentum transfer on the parameters of the laser
pulse is rather simple since the only involved parameter is $ T_{\rm ru}+2T_{\rm p}+T_{\rm rd}$.
In fact the dependence as a function of $F$ and the parameters of the laser pulse can be
encapsulated in a single effective parameter
$$
f = \frac{F}{2\sqrt{2}}\sqrt{T_{\rm ru}+2T_{\rm p}+T_{\rm rd}}, 
$$ 
so that for a fixed value of $f$, the formation probability no longer depends on the parameters of the laser pulse. 
Using the momentum transfer \ref{momentumTransfer}, the energy at the end of the laser pulse is 
\begin{equation}
\label{eq:Ef}
E_f=E_0+\frac{P_R^0 \Delta P_R}{\mu}+\frac{(\Delta P_R)^2}{2\mu}.
\end{equation}
According to Eq.~\ref{eq:Ef}, there is formation if $E_f<0$. 
Since $\Delta P_R$ is negative, the final energy $E_f$ can only be negative
(i.e., resulting in a formation) if $P_R^0$ is positive. This is a necessary but not a sufficient condition. 
If $F$ is too small, the final energy remains positive (and close to $E_0$) since the negative term
is insufficient to compensate for $E_0$, so there is no possibility for formation. If $F$ is too large, the dominant term in Eq.~\ref{eq:Ef} is $(\Delta P_R)^2/(2\mu)$ which is positive, therefore
resulting in a positive final energy and there is no formation. This qualitatively gives the
explanation for the increase of
the formation probability for small $F$ and the decrease for large $F$.

\medskip
In order to be more quantitative, we consider Eq.~\ref{eq:Ef} for $P_R^0 > 0$ as a general
 function $E(R)$ in the variable $R$ and which depends on the parameter $f$,
\begin{equation}
\label{eq:Phi}
E(R)=E_0-f^2 \sqrt{\frac{2}{\mu} \left(E_0+\frac{b_6}{R^6}  \right)} \ \left(\frac{2 d_2}{R^3}+\frac{3 d_3}{R^4}
\right)
+\frac{f^4}{2\mu}\left(\frac{2 d_2}{R^3}+\frac{3 d_3}{R^4}\right)^2. 
\end{equation}
When $R\rightarrow \infty$, $E(R)$ tends to $E_0$ and when $R \rightarrow 0$, $E(R)$ tends
to $+\infty$. The function $E(R)$ has two roots $R_1(f)$ and $R_2(f)$ such
that $R_1(f)<R_2(f)$. Because the function $E(R)$ is negative between these two roots,
if the interatomic distance 
is in the region where $E(R)$ is negative, e.g., between the roots $R_1(f)$ and
$R_2(f)$, then there is formation.
\begin{figure}
 \includegraphics[width=0.4\textwidth]{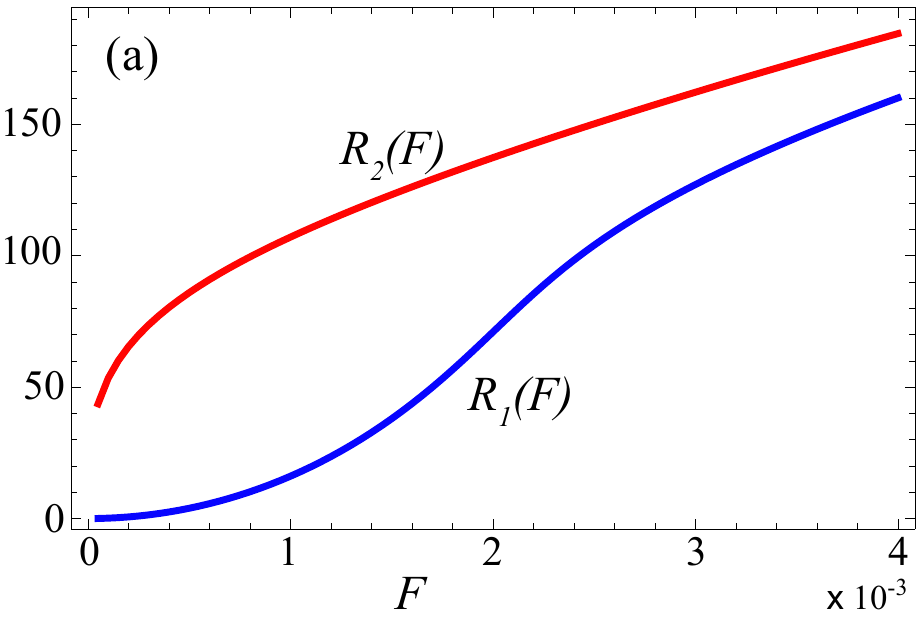} \qquad \includegraphics[width=0.395\textwidth]{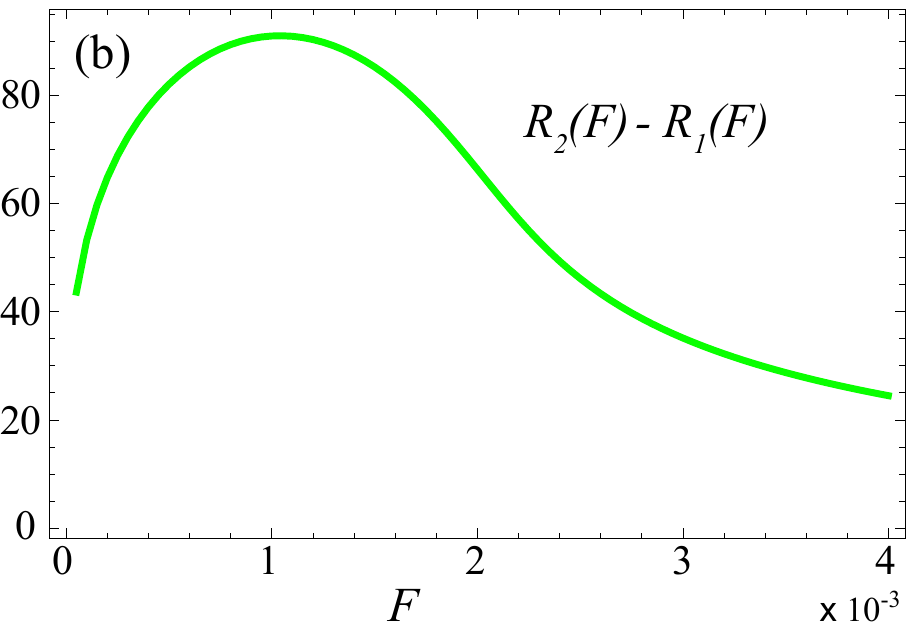}
 \caption{\label{fig:roots} a) Evolution as a function of $F$ of the roots $R_1$ and $R_2$
 of $E(R)$ given by Eq.~\ref{eq:Phi}.
 b) Evolution of $R_2 -R_1$ as a function of $F$.
 The parameters of the pulse are $T_{\rm ru}=15~\mbox{ps}$,
 $T_{\rm p}=70~\mbox{ps}$ and $T_{\rm rd}=15~\mbox{ps}$.}%
 \end{figure}
On Fig.~\ref{fig:roots}, are shown the evolutions of $R_{1,2}(F)$ and $R_2(F)-R_1(F)$ as a function
of $F$. We notice that the distance $R_2-R_1$ first increases with $F$ until $F\approx 10^{-3}~\mbox{a.u.}$ and then decreases. This behavior mirrors the increase and decrease of the
formation probability as a function of $F$.

\medskip
In Appendix~\ref{app:exp}, we derive some approximate expansions for the zeros of $E(R)$ and deduce two expansions for $R_2-R_1$, one for small values of $F$ and one for larger values of $F$. In a nutshell, these expansions lead to the following behaviors: for small $F$, the formation probability increases as $F^{2/7}$ and for large $F$, it roughly decreases with $F$ as $1/F$. More specifically, we have
\begin{eqnarray}
&& R_2(f)-R_1(f) \approx \left(\frac{3d_3\sqrt{2b_6}}{E_0\sqrt{\mu}}\right)^{1/7}f^{2/7} \qquad \mbox{ for } f\ll 1,\\
&& R_2(f)-R_1(f) \approx \frac{b_6^{1/2}(2\mu)^{1/4}}{2 \sqrt{3 d_3} E_0^{1/4} f}-\frac{b_6^{1/2}d_2(2\mu)^{1/8}}{4E_0^{3/8}(3d_3)^{5/4}f^{1/2}} \qquad \mbox{ for } f\gg 1.
\end{eqnarray}

\begin{figure}
 \includegraphics[width=0.4\textwidth]{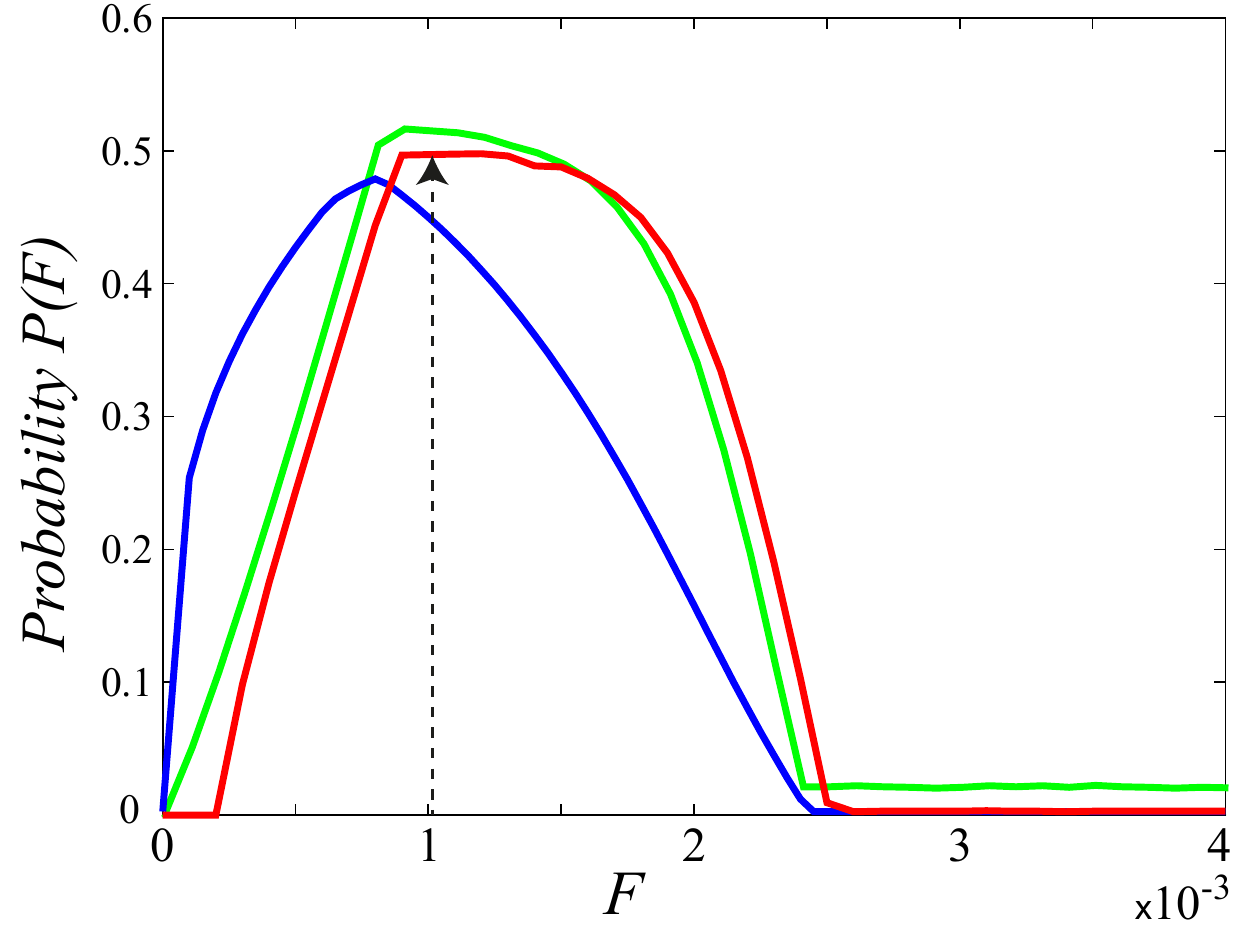}
 \caption{Formation probability given by Eq.~\ref{eq:ProbaFormula}
 as a function of $F$ (blue line). For completeness, the
 formation probability as a function of $F$ obtained from
 the long-range Hamiltonian~\ref{eq:Ham_s} (red line) 
 and from the full Hamiltonian \ref{eq:Ham2D_1} (green line) are also shown.
 The black vertical dashed arrow is located at the
 value $F\approx 0.00107$ a.u. 
given by Eq.~\ref{eq:fmax}. For this value of $F$, it is expected to find
the maximum of the formation probability.
 The parameters of the pulse are $T_{\rm ru}=15~\mbox{ps}$,
 $T_{\rm p}=70~\mbox{ps}$ and $T_{\rm rd}=15~\mbox{ps}$.
  The initial energy of the trajectories is $E_0=3\times 10^{-9}$ a.u.}
\label{fig:formation2D_formula}
 \end{figure}

\medskip\noindent
Naturally, for an ensemble of values of $R_0$ between $R_{\rm min}$ and $R_{\rm max}$, we consider
the overlap between the  intervals $[R_{\rm min}, R_{\rm max}]$ and $[R_1(f), R_2(f)]$, so that
an approximation of the formation probability is given by
\begin{equation}
P(f)=\frac{{\rm min}(R_{\rm max},R_2(f))-{\rm max}(R_{\rm min},R_1(f))}{2(R_{\rm max}-R_{\rm min})},
\label{eq:ProbaFormula}
\end{equation}
if $R_1(f)\leq R_{\rm max}$ and $R_2(f)\geq R_{\rm min}$, otherwise the probability is zero since there is no overlap between the available values of $R_0$ and the values of $R$ leading to a negative energy. 
The coefficient $1/2$ in the probability expression~\ref{eq:ProbaFormula}
comes from the fact that for a given $R$, there are two possible initial values for $P_R^0$, one
positive (and possibly leading to formation) and another one negative
(not leading to formation) with the same energy $E_0$. The blue
curve on Fig.~\ref{fig:formation2D_formula} is the formation probability obtained using the numerical
computation of the roots of $E(R)$ and using Eq.~\ref{eq:ProbaFormula}.
The agreement with the numerical integration
of the trajectories for the simplified Hamiltonian~\ref{eq:Ham_s} as well as with
the full one-dimensional Hamiltonian~\ref{eq:Ham2D_1} is very good, validating
the assumptions on the dynamics of the trajectories leading to the approximation \ref{eq:ProbaFormula}
for the formation probability.   
The main reason for the rather good quantitative agreement 
is that, in the interval $[R_{\rm min}, R_{\rm max}]$, a
large portion of the initial values of $R$ are large and the approximations performed to derive
Eq.~\ref{eq:ProbaFormula} are valid.

Three parameters emerge as most influential in the formation probability. All of them are
related to the long-range behavior of the dimer. One is related to the dimer potential
(behavior as $1/R^6$) and two are linked with the parallel polarizability
(behaviors as $1/R^2$ and $1/R^3$). It should be noticed that the term
in $1/R^6$ in the potential $\varepsilon(R)$ is absolutely essential to
ensure the existence of the two roots of $E(R)$.

In Appendix~\ref{app:exp} we also provide an approximate expression for the value of the electric
field amplitude where a maximum of formation is expected and it is given by
\begin{equation}
\label{eq:fmax}
F\approx \frac{2\sqrt{2}}{\sqrt{T_{\rm ru}+2T_{\rm p}+T_{\rm rd}}}. 
\end{equation}
\noindent
For a laser pulse with parameters $T_{\rm ru}=15~\mbox{ps}$,
 $T_{\rm p}=70~\mbox{ps}$ and $T_{\rm rd}=15~\mbox{ps}$, according to Eq.~\ref{eq:fmax},
 the maximum of formation is expected at $F\approx 0.00107$ a.u.
 As we can observe in Fig.~\ref{fig:formation2D_formula}, this value lies in
 the neighborhood of the values of $F$ where the computed formation probability is maximum.

In addition,  we have shown in Appendix~\ref{app:exp} the rather small dependence of
the formation probability with respect to the initial energy of the system
(or equivalently, to its temperature).

\section*{Conclusion}

The classical study carried out in this paper shows the feasibility of
using an intense linearly polarized laser field to drive the association of
Rb and Cs cold atoms to create a dimer in its ground state.
Interestingly, from our numerical calculations of the evolution of the formation probability
as a function of the electric field strength of the laser, we find that the formation probability first
increases and then decreases with increasing laser field intensity.
In order to explain this surprising behavior of the formation probability, we use nonlinear dynamics
and we show that
the main element responsible for the formation of RbCs is a rather small change in the
radial momentum $P_R$ induced by the laser pulse through its interaction with
the molecular polarizability. 
This change of radial momentum is so small that it is not sufficient
to induce changes in the positions of the atoms on the short-time scale of the laser pulse.
However it is sufficient to allow the formation of RbCs dimers.
Furthermore, the behavior of the formation probability
reflects the long-range behavior of the dimer.  The deep impact of the long-range behavior 
of the molecule in the formation mechanism allows us
to reduce the dynamics to a one dimensional radial Hamiltonian where only the long-range terms of
the potential are taken into account.
With this simplified Hamiltonian, we explained why initially positive
momentum leads to higher formation, and why an initially too short or too large interatomic
distance (i.e., shorter than $R_1(f)$ or larger than $R_2(f)$) does not lead to formation.
Moreover, from these observations and using that one dimensional Hamiltonian, we have derived
the approximate expression \ref{eq:ProbaFormula} for the
formation probability which highlights the role of the relevant parameters of laser
pulse and of the interaction potential which lead to the shaping of the formation
probability. In particular, such an expression might be helpful to control the formation probability
by adjusting the parameters of the laser field.
Finally, a quantum extension of our classical approach to the driven formation of cold dimers 
is of immediate interest in order to predict
the quantum association rate which could be compared to experiments. Work along this
line is now in progress.

\appendix

\section{Approximate expressions for the zeros of the function $E(R)$}
\label{app:exp}

In order to obtain the asymptotic behaviors of the zeros of Eq.~\ref{eq:Phi}
and hence of the formation probability, we rewrite $E(R)$ as
$$
E(R)=\frac{1}{2}\left[X-\sqrt{2\left(E_0+\frac{b_6}{R^6}\right)} 
\right]^2-\frac{b_6}{R^6},
$$
where 
$$
X=\frac{f^2}{\sqrt{\mu}}\left(\frac{2d_2}{R^3}+\frac{3d_3}{R^4} \right).
$$
The zeros of $E(R)$ satisfy
\begin{equation}
\label{eq:branches}
X_\pm=\sqrt{2\left(E_0+\frac{b_6}{R^6}\right)} \pm\sqrt{\frac{2b_6}{R^6}}.
\end{equation}
The above equation corresponds to two implicit equations for $R_1$ and $R_2$. 
The branch with $X_+$ corresponds to $R_1$ and the one with $X_-$ to $R_2$. When
$f$ tends to zero, the two solutions $R_1$ and $R_2$ converge to zero. Using an expansion of Eq.~\ref{eq:branches} around $R=0$, we obtain the asymptotic behaviors
\begin{eqnarray}
&& R_1(f)\approx \frac{3d_3}{2\sqrt{2\mu b_6}}f^2,\\
&& R_2(f)\approx \left(\frac{3d_3\sqrt{2b_6}}{E_0\sqrt{\mu}}\right)^{1/7}f^{2/7}. \label{eq:Bex1}
\end{eqnarray}
As a consequence, if $[R_1,R_2]\subset [R_{\rm min} R_{\rm max}]$,  the formation probability increases as $f^{2/7}$. It is worth noticing that there is a very slight dependence on the initial
energy (i.e., on the temperature T of the gas) since the approximate formation probability behaves as $T^{-1/7}$. 

For large values of $f$, the two roots $R_1(f)$ and $R_2(f)$ tend to infinity with the same asymptotic behavior given by $R_0(f)$ solution of
$$
\frac{f^2}{\sqrt{\mu}}\left(\frac{2d_2}{R^3}+\frac{3d_3}{R^4} \right)=\sqrt{2 E_0}.
$$
An explicit solution of $R_0(f)$ can be obtained since it is a solution of a quartic polynomial. However this expression is not very helpful. 
An expansion of the solution is given by
$$
R_0(f)=3^{1/4}d_3^{1/4}\frac{\sqrt{f}}{(2\mu E_0)^{1/8}}+
\frac{d_2}{2\sqrt{3d_3}}\frac{f}{(2\mu E_0)^{1/4}}+ O(f^{5/4}).
$$
The two roots $R_1(f)$ and $R_2(f)$ tend to $R_0(f)$ as $f$ increases, and
the distance between the two roots decreases as
\begin{equation}
\label{eq:Bex2}
R_2(f)-R_1(f)\approx \frac{b_6^{1/2}(2\mu)^{1/4}}{2 \sqrt{3 d_3} E_0^{1/4} f}-\frac{b_6^{1/2}d_2(2\mu)^{1/8}}{4E_0^{3/8}(3d_3)^{5/4}f^{1/2}}.  
\end{equation}
Given the values of the coefficients, we expect the formation probability
to decrease as $f$ increases. The leading behavior is proportional to $f^{-1}$
but the second term is of the same order, so it needs to be taken into account for
a more quantitative agreement (see Fig.~\ref{fig:roots}). We notice the strong
dependence of the formation probability with one of the parameters of the
potential $\varepsilon (R)$, namely $b_6$, as well as the two main parameters of the
parallel polarizability, namely $d_2$ and $d_3$. In addition, there is a slight dependence
of the initial energy (or equivalently the temperature): it increases as the temperature
decreases. The leading behavior is $T^{-1/4}$. 
Using Eqs.~\ref{eq:Bex1}-\ref{eq:Bex2}, we obtain an approximate value of $F$ for the
expected maximum of $R_2-R_1$:
$$
f_*=\frac{b_6^{1/3}\mu^{1/4}}{2^{23/36}(3d_3)^{1/2}E_0^{1/12}}.
$$
In particular we notice the very small dependence of this value with the initial energy, i.e., the temperature of the gas. As a rule of thumb, $f_*\approx 1$, so the expected maximum for the formation probability is approximately obtained for
$$
F_*\approx \frac{2\sqrt{2}}{\sqrt{T_{\rm ru}+2T_{\rm p}+T_{\rm rd}}}.
$$ 

\section*{Acknowledgements}

J.M. acknowledges the Research Committee of the University of Antioquia 
(CODI), Medell\'{\i}n, Colombia, through the project CODI-251594 and the 
``Estrategia de Sos\-te\-n\-ibilidad'' del Grupo de F\'{\i}sica 
At\'omica y Molecular. J.P.S. acknowledges financial support by the Spanish project
MTM-2014-59433-C2-2-P (MINECO) and the 
hospitality of the Grupo de F\'{\i}sica At\'omica y 
Molecular during his stay in September of 2015 at the University of Antioquia, Medell\'{\i}n, Colombia.

\end{document}